\documentclass[conference]{IEEEtran}
\IEEEoverridecommandlockouts

\usepackage[colorlinks=true, linkcolor=blue, urlcolor=cyan, citecolor=red]{hyperref}

\usepackage{cite}
\usepackage{amsmath,amssymb,amsfonts}
\usepackage{algorithmic}
\usepackage{graphicx}
\usepackage{textcomp}
\usepackage{xcolor}
\def\BibTeX{{\rm B\kern-.05em{\sc i\kern-.025em b}\kern-.08em
    T\kern-.1667em\lower.7ex\hbox{E}\kern-.125emX}}

\usepackage{caption}
\usepackage{float}
\newfloat{cascade}{hbtp}{lop}
\floatname{cascade}{Cascade}
\usepackage{mdframed}
\newcommand{\cascadecomment}[1]{\textcolor{teal}{/* #1 */}}
\usepackage{xspace}
\usepackage{comment}
\usepackage{subcaption}
\usepackage{bbm}
\usepackage{url}

\newcommand{\zspace}[1]{}

\newfloat{mapping}{hbtp}{lop}
\floatname{mapping}{Mapping}
\usepackage{mdframed}

\newcommand{\name}{FuseMax\xspace}
\newcommand{\EDGE}{EDGE\xspace}
\newcommand{\edge}{EDGE\xspace}

\begin{document}

\title{\name: Leveraging Extended Einsums to Optimize Attention Accelerator Design
\thanks{This work was partially funded by NSF grants 
CNS-1954521,
CNS-1942888,
CNS-2154183,
CCF-8191902, and 
CCF-2217099; 
as well as by an Intel gift and a Microsoft Research PhD fellowship.}
}

\author{\IEEEauthorblockN{Nandeeka Nayak}
\IEEEauthorblockA{
\textit{University of California, Berkeley}\\
Berkeley, CA, USA \\
nandeeka@berkeley.edu}
\and
\IEEEauthorblockN{Xinrui Wu}
\IEEEauthorblockA{
\textit{Tsinghua University}\\
Beijing, China \\
xr-wu20@mails.tsinghua.edu.cn}
\and
\IEEEauthorblockN{Toluwanimi O. Odemuyiwa}
\IEEEauthorblockA{
\textit{University of California, Davis}\\
Davis, CA, USA \\
todemuyiwa@ucdavis.edu}
\and
\IEEEauthorblockN{$\;\;\;\;\;\;\;\;\;\;$}
\IEEEauthorblockA{}
\and
\IEEEauthorblockN{$\;\;\;\;\;$Michael Pellauer}
\IEEEauthorblockA{
\textit{$\;\;\;\;\;\;\;$NVIDIA}\\
$\;\;\;\;\;\;\;$Westford, MA, USA \\
$\;\;\;\;\;\;\;$mpellaer@nvidia.com}
\and
\IEEEauthorblockN{Joel S. Emer}
\IEEEauthorblockA{
\textit{Massachusetts Institute of Technology / NVIDIA}\\
Cambridge, MA, USA \\
jsemer@mit.edu}
\and
\IEEEauthorblockN{Christopher W. Fletcher}
\IEEEauthorblockA{
\textit{University of California, Berkeley}\\
Berkeley, CA, USA \\
cwfletcher@berkeley.edu}
}

\maketitle

\begin{abstract}
Attention for transformers is a critical workload that has recently received significant `attention' as a target for custom acceleration.
Yet, while prior work succeeds in reducing attention's memory-bandwidth requirements, it creates load imbalance between operators that comprise the attention computation (resulting in severe compute under-utilization) and requires on-chip memory that scales with sequence length (which is expected to grow over time).

This paper ameliorates these issues, enabling attention with nearly 100\% compute utilization, no off-chip memory traffic bottlenecks, and on-chip buffer size requirements that are independent of sequence length.
The main conceptual contribution is to use a recently proposed abstraction---the \emph{cascade of Einsums}---to describe, formalize, and taxonomize the space of attention algorithms that appear in the literature.
In particular, we show how Einsum cascades can be used to infer non-trivial lower bounds on the number of \emph{passes} a kernel must take through its input data, which has implications for either required on-chip buffer capacity or memory traffic.
We show how this notion can be used to meaningfully divide the space of attention algorithms into several categories and use these categories to inform our design process.

Based on the above characterization, we propose \name---a novel mapping and binding of attention onto a spatial array-style architecture.
On attention, in an iso-area comparison, \name achieves an average $6.7\times$ speedup over the prior state-of-the-art, FLAT, while using $79\%$ of the energy.
Similarly, on full end-to-end transformer inference, \name achieves an average $5.3\times$ speedup over FLAT using $83\%$ of the energy.

\end{abstract}

\section{Introduction}
\label{sec:intro}

Over the past few years, transformers~\cite{attention} have emerged as the model architecture of choice for a wide range of machine learning applications, from natural language processing~\cite{bert, trxl-xlm, gpt, t5} to computer vision~\cite{vit, swin} to speech recognition~\cite{wav2vec, hubert}.
This rise has been accompanied by a corresponding wave of proposals for accelerating transformers in both software~\cite{choi-et-al-2022, flashattention2, flashattention} and hardware~\cite{flat, tileflow}.

Fortunately, many of the layers (projections, fully connected layers, etc.) used by transformers look very similar to prior generations of machine learning models.
Their resource-intensive tensor products can be described and evaluated with existing tensor algebra accelerator modeling tools~\cite{maestro, teaal, timeloop}, and many of the other layers (e.g., layer normalization) have negligible impact on performance and can be safely ignored.

However, the attention layer~\cite{attention}---usually described as a matrix multiplication, a softmax, and then another matrix multiplication---does not fit into either of these boxes.
For example, the softmax is both memory intensive (featuring low algorithmic reuse) \emph{and} compute intensive (featuring exponentiation and division).
Furthermore, attention's characteristics 
preclude many ``free lunches'' often used to improve efficiency in other DNN models.
For example, because all tensors are a function of the model inputs, there is no opportunity to amortize memory access costs with an increased batch size.
Additionally, since none of the operands can be computed before the inputs are given,  compression/strength reduction techniques (e.g., quantization~\cite{gobo, olive}, sparsity~\cite{spatten, sanger, dota}, etc.) must be applied dynamically, leading to 
more complicated algorithms and hardware designs.

To illustrate the difficulty in accelerating attention, consider the state-of-the-art accelerator for attention: FLAT~\cite{flat}.
FLAT uses fusion to reduce attention memory bandwidth bottlenecks on a spatial architecture (e.g., a TPU~\cite{tpu}).
Specifically, FLAT maps attention's matrix multiplications to the 2D spatial array and softmax operations to a separate 1D array.
While FLAT's design does make attention compute bound, it becomes compute bottlenecked in the 1D array (the softmax), causing severe under utilization of the 2D array.
While one could add additional PEs to the 1D array, this results in corresponding area costs.

Making matters worse, FLAT requires that the entire vector over which the softmax is performed be buffered on chip.
This vector is proportional to the sequence length, which is growing rapidly with time (e.g., Google reports 10 million length sequences in research, which would require 100s of MegaBytes to buffer~\cite{google10m}).
When the vector/sequence length grows beyond allowable buffer capacity, FLAT is forced to spill, which contributes significantly to attention energy consumption and can even make attention memory-bandwidth bound.

\textbf{This paper.}
We address the above challenges by proposing a novel spatial architecture -- \emph{\name} -- to accelerate attention, with particular emphasis on removing bottlenecks imposed by the softmax.
Our architecture addresses all of the aforementioned issues associated with FLAT.
Namely: 
\begin{itemize}
\item \name is compute bound, but provides almost 100\% utilization of both the 2D and 1D arrays throughout the attention layer, without adding additional PEs to the 1D array.
\item \name's on-chip memory requirements are invariant to sequence length and require no extra spills to memory regardless of sequence length.
\end{itemize}

The paper's technical core is split into three parts.

First, Section~\ref{sec:passes} demonstrates a novel analysis on kernels that uses the recently proposed \emph{cascade of Einsums} abstraction~\cite{teaal}.
In a nutshell, an Einsum defines an iteration space over tensors and what computation is done on and between tensors at each point in the iteration space.
A cascade of Einsums is a sequence of dependent Einsums that can be used to describe and specify a larger kernel.

While prior work~\cite{teaal, edge} provides a precise definition for Einsums, a major contribution in our work is to show how this definition can be leveraged to inform accelerator design.
Specifically, we recognize that the cascade makes explicit \emph{precisely} what dependencies there are between Einsums.
We show how this 
can be used to make non-trivial deductions about a kernel’s allowed fusion granularity
and algorithmic minimum per-tensor live footprint. 
The relationship between the live footprint and the buffer capacity, in turn, has implications for the required data movement.

In more detail, this analysis provides insight into the number of \emph{passes} an algorithm performs, i.e., the number of times a given element of an input must be revisited after visiting every other element of the input.
Normally, one strives to choose a dataflow that exploits maximal reuse in a given element (or tile of elements) to avoid having frequently reload it.
However, some algorithms preclude this strategy.
In this work, we describe how to count the number of passes a cascade requires and present two methods for reducing the number of passes.
In general, fewer passes is preferable; although, interestingly, we find that decreasing the number of passes can increase the required compute.
Given that an Einsum cascade is mapping/scheduling agnostic, this analysis provides insight given any possible scheduling of the cascade onto hardware.

Next, Section~\ref{sec:einsums} applies the cascade of Einsums abstraction to describe and formalize the attention kernel.
Using the notion of passes introduced in Section~\ref{sec:passes}, we taxonomize the space of numerically stable attention proposals that appear in the literature.
For example, in a na\"ive implementation of attention, one must traverse the entire softmax input to build the softmax denominator \emph{and only after that} can one revisit and scale each input (softmax numerator) by the denominator.
Because this analysis is performed on the cascade of Einsums, our lower bounds on passes hold for any choice of mapping, including applications of fusion.
For example, despite using fusion, FLAT employs a 3-pass cascade and its reliance on large on-chip buffering is a symptom of trying to avoid three passes-worth of DRAM traffic.
We, then, show how transforming the attention cascade reduces the number of passes required.

Additionally, we find that expressing attention as a cascade of Einsums reveals that optimizations that were previously conflated can actually be applied separately.
We specifically call out one that is used by 1-pass algorithms to eliminate the need for a second pass after the final softmax denominator has been calculated.
We recognize that this optimization has the added benefit of decreasing the required divisions, 
which is not only useful for but can be applied to 2- and 3-pass cascades as well.

Finally, Section~\ref{sec:mapping} uses the insights from Section~\ref{sec:einsums} as a starting point to develop a novel mapping and binding for attention that can be lowered to a spatial architecture.
We call our architecture \name. 
\name adopts the 1-pass attention cascade used in FlashAttention-2~\cite{flashattention2}.
However, despite using the cascade from FlashAttention-2, binding this cascade to a spatial architecture is non-trivial.
In particular, FlashAttention-2 binds the cascade onto a GPU, an architecture that features homogeneous PEs, each with relatively large per-PE storage, and expensive inter-PE communication.
Spatial architectures feature opposite characteristics: heterogeneous PEs, each with smaller per-PE storage, and cheap (but restricted) inter-PE communication.
Specifically, the networks that connect the PEs within the 2D array allow efficient communication primarily between neighbors.
We overcome these differences and demonstrate a novel mapping and binding for the 1-pass cascade that achieves high utilization for entire transformer layers.
Our architecture requires only minimal changes to a standard spatial architecture and is performance/energy robust to long sequence lengths (e.g., 1M tokens and beyond).

To summarize, we make the following contributions:
\begin{itemize}
\item We show how cascades of Einsums can be used to inform accelerator design, both in terms of reasoning about compute requirements and per-tensor live footprints.
We formalize lower bounds on the number of passes a cascade imposes given any possible mapping of the cascade onto hardware.  
\item We use cascades of Einsums, and the observation about pass lower bounds, to provide a taxonomy and precise specification of numerically stable attention algorithms in the literature.
Orthogonally, we show how previously entangled attention optimizations can be applied across attention algorithms.
\item We propose a novel mapping and binding for attention for a spatial architecture---which we call \name---that achieves high utilization for both 2D and 1D array PEs, and has memory traffic requirements that are independent of sequence length.
\item We evaluate \name on BERT~\cite{bert}, TrXL~\cite{trxl-xlm}, T5~\cite{t5}, and XLM~\cite{trxl-xlm} and demonstrate a $6.7\times$ speedup on attention with $79\%$ of the energy and a $5.3\times$ speedup on full end-to-end inference with $83\%$ of the energy relative to FLAT.
\end{itemize}
\section{Background}\label{sec:background}
In this section, we describe the concepts and terminology used in the remainder of the paper.

\subsection{Tensors}\label{sec:background:tensors}
This paper focuses on algebraic computations on tensors, where a tensor is a multidimensional array.  
A tensor's \emph{rank} refers to a specific dimension of the tensor, while the tensor's \emph{shape} is the set of valid coordinates for each of the tensor's ranks. 
We use the notation $N$-tensor to denote a tensor with $N$ ranks, where a 0-tensor is a scalar, a 1-tensor is a vector, a 2-tensor is a matrix, etc.

We adopt the format-agnostic \emph{fibertree} abstraction of tensors, where a tensor is represented as a tree of fibers, as detailed in prior work~\cite{Wu:2022:SAA,sze:2020:epo, Odemuyiwa:2023:ASD, sam, Wu:2023:HEF, Pellauer:2023:SOS, teaal,unified_convolution_framework}, using the specific version described in TeAAL~\cite[Section 2.1]{teaal}. 
In this abstraction, a \emph{fiber} consists of the set of coordinates for a given rank with common coordinates for all higher-level ranks. Each coordinate is coupled with a \emph{payload}.
The payload may contain a reference to a fiber in the next lower rank, or to a leaf data \emph{value}.

\subsection{Traditional Einsums}\label{sec:background:einsums}
An Einsum expression defines a computation on a set of tensor operands using an iteration space that specifies the set of points where the computations are performed~\cite{teaal, edge}.
For example, we describe matrix-matrix multiplication (GEMM) with the following Einsum:
\begin{align}
Z_{m, n} = A_{k, m} \times B_{k, n}
\end{align}
where $A$ and $B$ are input 2-tensors of shape $K \times M$ and $K \times N$, respectively.
$Z$ is an output 2-tensor with shape $M \times N$. 
Throughout this paper, we use the same symbol for both the shape and \emph{name} of a rank (e.g., rank $K$ in $A$ has a shape of $K$).

The \emph{iteration space} of this Einsum is $[0, K) \times [0, M) \times [0, N)$.
An evaluation of this Einsum must: (1) walk every $(k, m, n)$ point in the iteration space; and, at each point (2) project into the \emph{data space} of all input tensors, (3) multiply the corresponding data values, and (4) place the result at the corresponding data point in $Z$. 
If a value already exists at an $(m, n)$ point in $Z$ (due to computation at the same $(m, n)$ point for a different $k$ in the iteration space), reduce the two values together using addition. Note that the Einsum specifies \emph{what} to compute; it does not indicate the order in which one walks the iteration space.
These aspects are left to the \emph{mapping}~\cite{eyeriss,timeloop,teaal}.

We also note that we can view the iteration space itself as a tensor.
In the example above, this tensor has shape $K \times M \times N$.
Therefore, we define a special fibertree---called the iteration space fibertree or \emph{is-fibertree}---that is the fibertree representation of this iteration space tensor.

\subsection{Extended Einsums}\label{sec:background:edge}
Traditional Einsums sufficiently express standard tensor algebra, including those supported in Basic Linear Algebra Subprograms (BLAS)~\cite{Lawson:1979:BLA, Duff:2002:AOS} and tensor network contractions~\cite{Ran:2020:TNC}. 
However, they cannot handle more complex computations.
The recently proposed Extended General Einsums notation (\EDGE)~\cite{edge}, extends Einsums to handle graph algorithm computations. 
We find this abstraction useful for also expressing complex tensor algebra computations and use its notation throughout the paper.
We now briefly summarize the portions of \edge{} that we leverage.

\subsubsection{User-Defined Computations}\label{sec:background:udf}
\EDGE separates computations into three ``actions'': map ($\bigwedge$), reduce ($\bigvee$), and populate ($=$)~\cite{edge}.
Map specifies the pair-wise computation between the shared ranks of two tensors, reduce specifies the computation for the reduction step of an Einsum, and default populate ($=$) places a computed value from the right-hand side (RHS) of the Einsum to its location on the left-hand side (LHS). 

Each map and reduce action contains two operations: merge and compute. 
Compute defines the operation to apply between two data values, and can be \emph{any} user-defined function. 
Merge specifies which regions of the iteration space to touch; execution will not need to access the data space corresponding to culled points.
Together, merge and compute precisely define the computations in an Einsum. 
Common merge operations include intersection ($\cap$), which touches points with non-zero values in \emph{both} operands; and union ($\cup$), which touches points where at least one of the operands is non-zero.

The full \edge{} specification for GEMM is then:
\begin{align}\label{eq:edge-gemm}
Z_{m, n} = A_{k, m} \cdot B_{k, n} :: \bigwedge_k \times(\cap) \bigvee_k +(\cup),
\end{align}
where $\bigwedge_k$ specifies a map action between $A$ and $B$ on the $k$ rank and the intersection merge operator ($\cap$) culls $k$ points where at least one operand is 
zero.
The compute operator ($\times$) multiplies the data values of coordinates surviving intersection. 
The reduce action ($\bigvee_k$) on the $k$ rank gathers all non-empty points in the $k$ rank and reduces them using addition ($+$). 

In this work, we use three user-defined computations: 
\begin{enumerate}
    \item Maximum ($\max(\cup)$) takes the maximum of two values.
    Suppose we have the following expression: $Z_m = A_m \cdot B_m :: \bigwedge_m \max(\cup)$.
    The union merge operator ($\cup$) filters out any $m$ coordinates where both operands contain $0$ (and places 0 in the output). 
    The $\max$ compute operator then returns the maximum of the two operands.
    \item Divide ($\div(\leftarrow)$) divides two data values. 
    Given the following expression, $Z_m = A_m \cdot B_m :: \bigwedge_m \div(\leftarrow)$,  the merge operator ($\leftarrow$) only touches $m$ points where there is a non-zero value in the $B$ operand (see~\cite[Appendix]{edge}), and the compute operator divides the data value in $A$ with the data value in $B$. 
    \item Subtraction and Exponentiation:
    To apply the exponential to an expression that subtracts two tensors, we use the following expression: $Z_m = A_m \cdot B_m :: \bigwedge_m \text{sub-then-exp}(\mathbbm{1})$. The user-defined operator ($\text{sub-then-exp}$) performs $A_m$ minus $B_m$ then applies the exponential to the result. The merge operator, $\mathbbm{1}$, is \edge's ``pass-through'' operator, which touches all $m$ points in the iteration space.
\end{enumerate}

In addition to map and reduce, \edge enables the expression of user-defined \emph{unary} operations on tensors.
For example, we can express the application of the non-linear, sigmoid function ($\sigma$) on each element of a tensor $A$ as $Z_m = \sigma(A_m)$.

\subsubsection{Shorthand Notation}\label{sec:background:shorthand}
Throughout this paper, we take advantage of \edge's shorthand notation~\cite{edge} in the following ways:
\begin{itemize}
 \item We drop all reduce actions that consist of add and union in the compute and merge operator, respectively ($\bigvee +(\cup)$). 
 Thus, $Z_m = A_{k, m} :: \bigvee_k +(\cup)$ becomes $Z_m = A_{k, m}$.
 \item We express all map actions using infix notation; that is, $A_{k, m} \cdot B_{k, n} :: \bigwedge_k \times(\cap)$ becomes $A_{k,m} \times B_{k, n}$. 
 \item When $\max$ is part of a map action ($A_{m} \cdot B_{m} :: \bigwedge_m \max(\cup)$), we replace it with the following shorthand: $\max(A_m, B_m)$.
 \item When $\div$ is part of a map action ($A_{m} \cdot B_{m} :: \bigwedge_m \div(\leftarrow)$), we replace it with the following: $A_m/B_m$.
 \item When sub-then-exp is part of the map action ($A_m \cdot B_m :: \bigwedge_m \text{sub-then-exp}(\mathbbm{1})$), we replace it with the shorthand
    $e^{A_m - B_m}$.
 \item We can express rank variable expressions with only one valid coordinate (e.g., $S_{i:i = 2}$) using just the coordinate (in this case, $S_2$).
\end{itemize}

\subsubsection{Filtering Rank Expressions}
\label{sec:background:indexfilter}

\edge also enables expressing Einsums that touch only a subset of the data space of their constituent tensors.
For example, we may express the prefix sum of a tensor $A_k$ with the following Einsum:
\begin{align*}
S_{i + 1} = A_{k:k \leq i}
\end{align*}
For each coordinate $i$, $S_{i + 1}$ is built by reducing together the subset of $A$ whose coordinates are $\leq i$.
Note that this definition of prefix sum computes the entire sum for a given $i$ without iteratively reusing the previous sum.

\subsubsection{Expressing Iterative Computations}\label{sec:background:iterativerank}
EDGE expresses recursion and iteration through generative/iterative ranks.
We use the term \emph{standard} ranks to differentiate non-iterative ranks from iterative ranks. 
We can express the iterative prefix sum as follows:
\vspace{-0.1in}
\begin{align}
S_{i + 1} &= S_i + A_i \label{eq:iter:s} \\
&\diamond: i \geq K \label{eq:iter:sstop}
\end{align}
Here, $S$ is a tensor with the iterative rank, $I$, ranging from $0$ to $K$ (inclusive).
Statement~\ref {eq:iter:sstop} indicates the stopping condition for the iterative expression (when $i$ is greater than or equal to $K$).

\subsubsection{Cascades of Einsums}\label{sec:background:cascades}
TeAAL~\cite{teaal} introduces the concept of \emph{cascades} of Einsums, which expresses directed acyclic graphs (DAGs) of Einsum expressions as a sequence of sub-Einsums. 
One can view the unrolled iterative expression in Einsum~\ref{eq:iter:s} as a cascade:
\vspace{-0.1in}
\begin{align*}
S_{1} &= S_0 + A_0 \\
S_{2} &= S_1 + A_1 \\
...& \\
S_{K} &= S_{K - 1} + A_K
\end{align*}
\vspace{-0.1in}

Finally, we use the \edge \emph{Initialization} label to specify computations that initialize tensors, which occur once. We use the \edge \emph{Extended Einsum(s)} label to specify the computation that occurs on each iteration of a cascade of Einsums~\cite{edge}.
For example, see (Einsum) Cascade~\ref{fig:1pass}. 

\subsection{Mapping and Binding}\label{sec:background:mapping}
While the cascade of Einsums specifies what computation is required, the \emph{mapping} and \emph{binding} describe how it should occur
~\cite{eyeriss, timeloop, teaal, sze:2020:epo}.
We use the concept of \emph{logical tasks} to define these terms.
A logical task is a grouping of points in the iteration spaces of all Einsums.
Tasks are defined such that each point in the iteration spaces is assigned to exactly one task.
Logical tasks can be as small as a single point or as large as an entire iteration space.
In the final schedule, each logical task must be assigned to exactly one compute unit that finishes the given task before moving onto the next task.

The mapping, therefore, describes a \emph{task graph}, a directed, acyclic graph whose nodes are logical tasks and edges are dependencies between the tasks.
Mapping specifications typically include aspects such as loop order, partitioning, and work scheduling (sequential vs. parallel operations)~\cite{teaal}.
Thus, the dependencies in the task graph can be true dependencies (enforced by the cascade) or additional ordering constraints imposed by the mapping specification.

The binding describes how the tasks are bound to the actual hardware, including which compute unit each task is associated with, when that task will be executed, and where the inputs and outputs are stored in the memory hierarchy.
This binding must obey the dependencies present in the task graph and the physical limitations of the architecture but is otherwise unconstrained.

\subsection{Tensor Algebra Accelerators}\label{sec:background:accelerators}
In recent years, the popularity of domain-specific tensor algebra accelerators has increased. 
A typical accelerator based on a spatial architecture consists of off-chip main memory, an on-chip shared global buffer, various scratchpads, and a 1D and/or 2D processing engine (PE) array where each PE contains compute units~\cite{tileflow, flat, tpu, Odemuyiwa:2023:ASD, eyeriss}. 
This design minimizes memory transfer latency while maximizing compute utilization~\cite{dadiannao,diannao,eyeriss,tpu,chen:2020:SAA}.
Various tools enable the quick modeling and design space exploration of tensor algebra accelerators, including Timeloop~\cite{timeloop} and Accelergy~\cite{accelergy}, GAMMA~\cite{gamma}, and DOSA~\cite{Hong:2023:DDM}.
\section{Passes Performed by a Cascade of Einsums}
\label{sec:passes}

Our first contribution is to demonstrate a novel analysis that can be applied to a cascade of Einsums.
The key insight is that cascades of Einsums provide a precise description of the iteration space for each Einsum and the data space for each constituent tensor, enabling us to derive the algorithmic minimum live footprint for each tensor, with implications for the allowed fusion schedules and required buffer capacity/memory traffic.
Because this analysis relies only on the cascade of Einsums, it holds for any choice of mapping and binding.

\subsection{Calculating the Number of Passes}

We will apply our analysis to attention in Section~\ref{sec:einsums}.
To illustrate ideas, we first start with a simple pedagogical example, shown in Cascade~\ref{fig:pass:example}.

\vspace{-0.1in}
\begin{cascade}[H] 
    \caption{\label{fig:pass:example} An example 2-pass cascade.}
    \begin{mdframed}
\begin{align}
Y &= A_k \times B_k \label{eq:example:k:y}\\
Z &= Y \times A_k \label{eq:example:k:z}
\end{align}
    \end{mdframed}
\end{cascade}
\vspace{-0.1in}

Einsum~\ref{eq:example:k:y} performs a dot product between $A_k$ and $B_k$, and Einsum~\ref{eq:example:k:z} multiplies the first Einsum's result $Y$ by $A_k$ again to produce $Z$.
If we want to minimize data traffic of $A_k$, we need to choose a dataflow for each Einsum that keeps $A_k$ stationary and fuses the two Einsums together.
In other words, the dataflow must finish using the first element of $A_k$ before moving onto the next.
However, such a dataflow does not exist for this cascade.
Any implementation must visit \emph{every} element of $A_k$ to compute $Y$ before it can revisit \emph{any} element of $A_k$ to compute $Z$.

We define a \emph{pass} that a cascade performs over a particular fiber of a particular rank and tensor to be a traversal of every element of that fiber.
Each time an element \emph{must} be revisited \emph{after} visiting every other element of that fiber, there is an additional pass.
For example, Cascade~\ref{fig:pass:example} performs two passes over the $K$ rank of $A_k$.

Since an Einsum's iteration space can also be represented as a fibertree (i.e., an \emph{is-fibertree} -- see Section~\ref{sec:background:einsums}), we extend our definition of an iteration space for a cascade of Einsums by considering its iteration space to be the sequence of the is-fibertrees for each Einsum. 
Now, in a scenario where fibers for a particular rank exist in multiple is-fibertrees; in each, they project to the same tensor; and there is a dependency such that all of the elements of the earlier is-fibertree's fiber must be read before any element can be read again by the later is-fibertree (for all mappings of the cascade), we refer to that read-read sequence as creating an additional \emph{pass}.
When there is a sequence of $N$ such read-read dependencies, we say the cascade is an $(N+1)$-pass cascade.
For our example, Cascade~\ref{fig:pass:example} requires two passes of the $K$ rank.

\subsection{Implications of the Number of Passes}

The number of passes a cascade performs is relevant because it restricts possible fusion schedules. 
Einsums within a pass can be fused at will, producing and consuming a tile of the intermediate at a time.
Einsums in different passes cannot be fused. 
Revisiting Cascade~\ref{fig:pass:example}, Einsums~\ref{eq:example:k:y} and \ref{eq:example:k:z} cannot be fused on the $K$ rank.
Any implementation must visit all elements of the $K$ fiber of $A$ to produce $Y$ before it can visit any of the elements of that fiber to produce $Z$.

This analysis also provides a non-trivial lower bound on the tensors' live footprints. 
For example, the algorithmic minimum live footprint for tensor $A$ is a fiber of shape $K$.
In other words, an architecture must either have enough buffer space to hold an entire $K$ fiber of $A$ or spill and reload that fiber, incurring memory traffic proportional to the shape of $K$.
We note that this analysis is mapping independent. There is no dataflow for this cascade that enables a smaller live footprint.

\subsection{Reducing the Number of Passes via Reassociation}
\label{sec:passes:reassociation}

Given the restrictions that multi-pass cascades place on the allowed dataflows and tensor live footprints, it can be beneficial to manipulate the cascade to reduce the number of passes required.
Crucially, these manipulations are functionally equivalent and only change how $Z$ is computed.
In this section, we will present two methods for doing so, though we leave a full analysis of the space of pass-reduction approaches to future work.

\subsubsection{Deferring the Multiplication by $Y$}
\label{sec:passes:reassociation:defer}

First, we recognize that, by the distributive property, Einsum~\ref{eq:example:k:z} can be factored to perform the reduction of $A_k$ first, before multiplying the result by $Y$.
Doing so, we get the following cascade:
\vspace{-0.1in}
\begin{cascade}[H] 
    \caption{\label{fig:pass:deferred} A reassociation of Cascade~\ref{fig:pass:example} that defers the $Y \times$ to compute $Z$ with 1-pass of the $K$ rank.}
    \begin{mdframed}
\begin{align}
Y &= A_k \times B_k \label{eq:deferred:y}\\
X &= A_k \label{eq:deferred:x} \\
Z &= Y \times X \label{eq:deferred:z}
\end{align}
    \end{mdframed}
\end{cascade}
\vspace{-0.1in}
Now, because there is no read-after-write dependency between Einsums~\ref{eq:deferred:y}~and~\ref{eq:deferred:x}, both Einsums can be included in the same pass.
In fact, because Einsum~\ref{eq:deferred:x} reduces away the $K$ rank, Cascade~\ref{fig:pass:deferred} is a 1-pass cascade with respect to this rank.
This reassociation actually provides a second benefit over Cascade~\ref{fig:pass:example}: Einsum~\ref{eq:deferred:z} now only requires one multiplication (as opposed to $K$ multiplications in Einsum~\ref{eq:example:k:z}).

\subsubsection{Iteratively Constructing $Y$ and $Z$}
\label{sec:passes:reassociation:iter}

\begin{cascade}[h] 
    \caption{\label{fig:pass:iter} A reassociation of Cascade~\ref{fig:pass:example} that iteratively constructs $Y$ and $Z$ with 1-pass of the $K$ rank.}
    \begin{mdframed}
Initialization:
\begin{align}
RY_{i:i=0} = 0 \\
RZ_{i:i=0} = 0
\end{align}
Extended Einsums:
\begin{align}
RY_{i + 1} &= RY_{i} + A_{i} \times B_{i} \label{eq:iter:ry}\\
RZ_{i + 1} &= RZ_i \times \frac{RY_{i + 1}}{RY_i} + RY_{i + 1} \times A_i \label{eq:iter:rz}\\
Z &= RZ_{K} \\
&\diamond: i \geq K \label{eq:iter:istop}
\end{align}
    \end{mdframed}
\end{cascade}

Alternatively, we can iteratively construct $Y$ and $Z$ as we perform the pass through $A_k$.
To do so, we will take a similar approach to the prefix-sum (see Sections~\ref{sec:background:indexfilter}-\ref{sec:background:iterativerank}) and build intermediate $Y$s and $Zs$.
\begin{align}
RY_{i + 1} &= A_{k:k \leq i} \times B_{k:k \leq i} \label{eq:filter:ry}\\
RZ_{i + 1} &= RY_{i + 1} \times A_{k:k \leq i}\label{eq:filter:rz}
\end{align}
Just like with the prefix sum, this version requires a lot of extra compute, but, because $Y = RY_K$ and therefore $Z = RZ_K$, the final result is the same. 

We remove this extra work by making the $I$ ranks of $RY_{i + 1}$ and $RZ_{i + 1}$ iterative.
This is shown in Cascade~\ref{fig:pass:iter}.
Iterative $RY_{i + 1}$ (Einsum~\ref{eq:iter:ry}) looks very similar to the iterative prefix-sum.
However, computing $RZ_{i + 1}$ is a little more complicated.

To derive the expression for $RZ_{i + 1}$, we start by introducing one more intermediate $S_i$, which is the prefix sum for $A_k$:
\begin{align}
S_{i} = A_{k:k \leq i - 1} \label{eq:pass:sum}
\end{align}
Now, we can combine Einsums~\ref{eq:filter:rz}~and~\ref{eq:pass:sum} to write $RZ_i$ in terms of this prefix-sum:
\begin{align}
RZ_{i} = RY_{i} \times S_i \label{eq:iter:rz_i}
\end{align}
Dividing both sides by $RY_{i}$, we derive an alternate definition for $S_i$:
\begin{align*}
S_i = \frac{RZ_i}{RY_i}
\end{align*}
$S_{i + 1}$ can also be written using this alternative definition:
\begin{align}
S_{i + 1} = \frac{RZ_i}{RY_i} + A_i \label{eq:iter:s_i+1}
\end{align}
We can combine Einsums~\ref{eq:iter:rz_i}~and~\ref{eq:iter:s_i+1} to compute $RZ_{i + 1}$ in terms of $RZ_i$ (i.e., iteratively):
\begin{align*}
RZ_{i + 1} = RY_{i + 1} \times \left(\frac{RZ_i}{RY_i} + A_i\right)
\end{align*}
Distributing $RY_{i + 1}$ and performing some reassociation, we get Einsum~\ref{eq:iter:rz}.

Cascade~\ref{fig:pass:iter} is also a 1-pass cascade, performing one pass of the $K$ rank of $A_k$ (indexed with the variable $i$) and iteratively building $RY_{i + 1}$ and $RZ_{i + 1}$.
Unfortunately, unlike Cascade~\ref{fig:pass:deferred}, Cascade~\ref{fig:pass:iter} does require extra compute over the original Cascade~\ref{fig:pass:example}.
However, memory bandwidth-limited workloads can afford to trade off extra compute for reduced memory traffic, and Cascade~\ref{fig:pass:iter} may still provide benefit.
\section{Taxonomizing Attention as Einsum Cascades}
\label{sec:einsums}

Our second contribution is to apply the cascade of Einsums abstraction and the notion of passes to transformer models to describe, taxonomize, and highlight trade-offs in the space of attention implementations.
This section first looks at the transformer model as a whole, identifying attention as an important kernel (Section~\ref{sec:background:transformers}).
We then give an overview of attention and a ``straightforward'' (but inefficient) algorithm for softmax by writing them as cascades of Einsums (Sections~\ref{sec:einsum:matmul}-\ref{sec:einsums:softmax}).
Finally, we show how optimizations to softmax can be described by modifying the cascades and provide a taxonomy of the space using the number of passes required by each cascade (Sections~\ref{sec:einsums:opt-softmax-compute}-\ref{sec:einsums:opt-softmax-traffic}).

\subsection{Transformers}
\label{sec:background:transformers}

\begin{figure}[t!]
    \centering
    \begin{subfigure}[t]{0.24\textwidth}
        \centering
        \includegraphics[width=\textwidth]{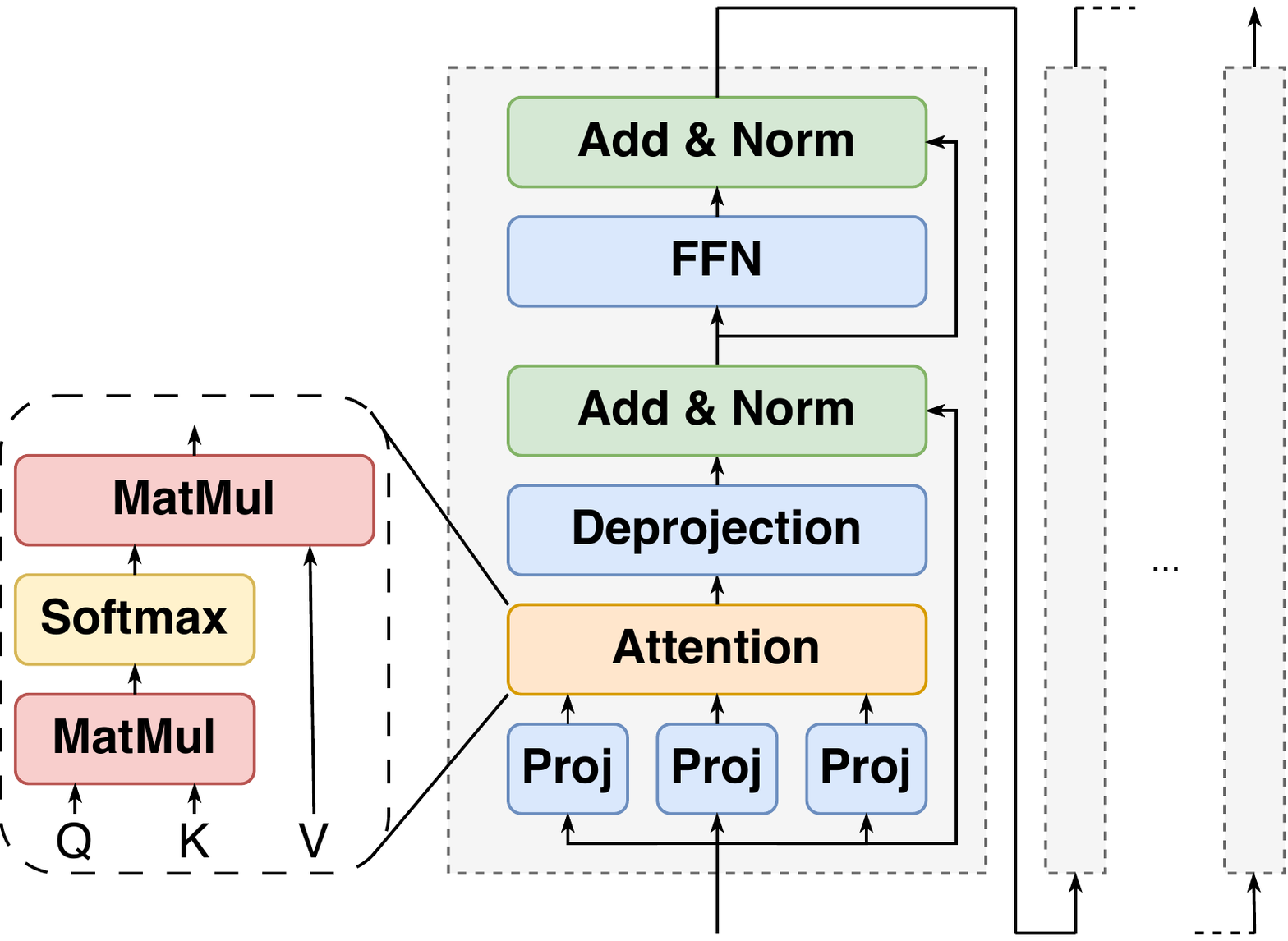}
        \caption{Encoder architecture}
        \label{fig:transformer:model}
    \end{subfigure}%
    ~ 
    \begin{subfigure}[t]{0.24\textwidth}
        \centering
        \includegraphics[width=\textwidth]{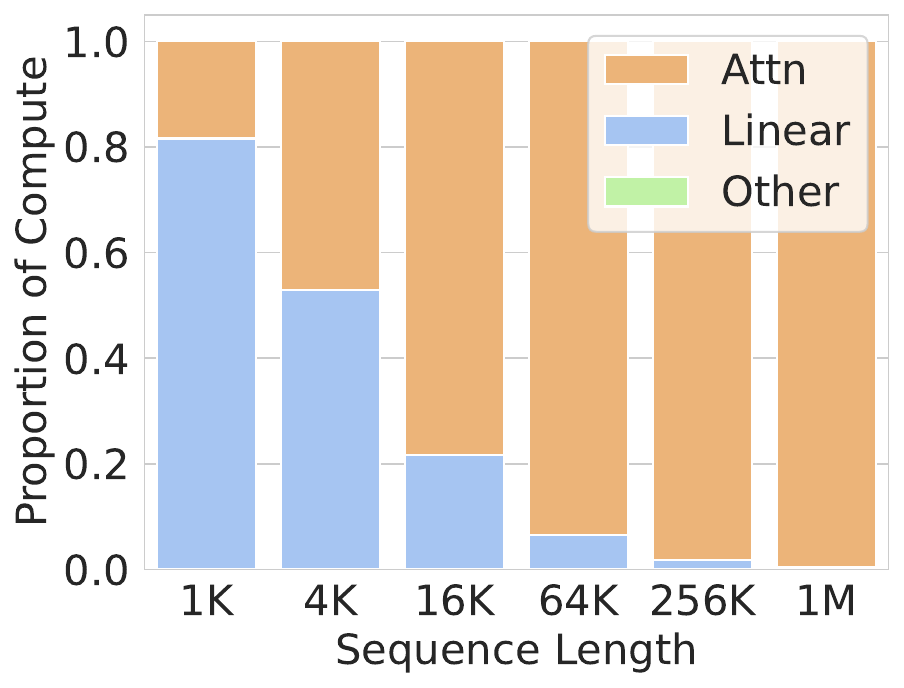}
        \caption{Required compute}
        \label{fig:transformer:ops}
    \end{subfigure}
    \caption{Overview of transformer encoder inference.}
    \label{fig:transformer:overview}
\vspace{-0.2in}
\end{figure}

Transformer models generally follow the architecture defined in \cite{attention}.
Our work, which addresses the impact of long sequence lengths during self-attention, focuses on the encoder architecture.\footnote{During the decoder phase, inference is severely bottlenecked on the memory traffic required to read the KV cache~\cite{kvquant}, and therefore the on-chip accelerator design has less impact on performance.}
Figure~\ref{fig:transformer:model} gives an overview.
The transformer first projects the input (by multiplying it by weight tensors) to form a \emph{query}, \emph{key}, and \emph{value}.
Self-attention is made up of three operations: a matrix multiplication of the query and key, a softmax on the result, and another matrix multiplication, which combines the softmax output with the value.
The attention output is then deprojected (again, multiplying by a weight tensor), normalized, passed through a two-layer feed-forward neural network (FFN), and normalized once more.

As the sequence length grows, the relative importance of the different operations changes. 
Figure~\ref{fig:transformer:ops} shows that at shorter sequence lengths, the \emph{weight-times-activation} ``linear" layers are a larger fraction of the total required compute, while at long sequence lengths, the attention operation dominates.
In all cases, the additional non-linearities (e.g., the normalization, the ReLU between the FFN layers, etc.) have negligible impact.
In the next section, we focus on describing attention more precisely, and use our analysis to understand prior work on efficient implementations.

\subsection{Redefining Attention's ``Matrix Multiplications''}
\label{sec:einsum:matmul}

In the original transformer paper~\cite{attention}, the kernel was described with the following equation:
\begin{align}
Attention(Q, K, V) = softmax\left(\frac{QK^T}{\sqrt{d_k}}\right)V
\end{align}

However, this equation says almost nothing about what the inputs $Q$, $K$, and $V$ look like or what iteration space needs to be traversed.
We clarify these points by rewriting the above
as a cascade of Einsums, with the exception of the softmax, whose cascade we will explore in Section~\ref{sec:einsums:softmax}.
The first step is to give each of the ranks names: $M$ and $P$ are the sequence lengths for $Q$ and $K$/$V$, respectively, and $E$ and $F$ are the embeddings for $Q$/$K$ and $V$, respectively.

\begin{align}
QK_{m, p} &= \frac{1}{\sqrt{E}} \times Q_{e, p} \times K_{e, m} \label{eq:attn:basic:qk} \\
A_{m, p} &= softmax(QK_{m, p}) \\
AV_{f, p} &= A_{m, p} \times V_{f, m} \label{eq:attn:basic:av}
\end{align}

Here, Einsums~\ref{eq:attn:basic:qk}\footnote{Einsums do not require the transpose, since this information is implicit in the indices.}\textsuperscript{,}\footnote{In Einsum~\ref{eq:attn:basic:qk}, we also substitute $E$ for $d_k$ following the notation defined in Section~\ref{sec:background:einsums}, where the shape of a rank is also its rank name.} and~\ref{eq:attn:basic:av} look like matrix multiplications.
Taking Einsum~\ref{eq:attn:basic:av} as an example, for each point in the iteration space $F \times M \times P$, we perform a multiplication using elements from two 2-tensors ($A_{m, p}$ and $V_{f, m}$) to produce a 2-tensor output ($AV_{f, p}$), which requires reducing across the inputs' shared rank $M$.
Einsums~\ref{eq:attn:basic:qk}-\ref{eq:attn:basic:av} can be modified to refer to the full batched, multi-head self attention~\cite{attention} by adding the batch ($B$) and head ($H$) ranks to all tensors.
This changes the characteristics of the kernel.
Adding the $B$ and $H$ ranks means that Einsums~\ref{eq:attn:basic:qk} and~\ref{eq:attn:basic:av} behave like many independent matrix multiplications instead of one monolithic matrix multiplication.
The challenges with attention, described in Section~\ref{sec:intro}, 
still follow clearly from this modification.
Because \emph{all} tensors contain a $B$ rank, the matrix multiplications are all unique to the specific batch's inputs.
Therefore, 
none of these tensors can be computed before the inputs are given, and there is no data sharing between the different
elements in the batch.
Hence, to simplify notation, we assume the presence of the $B$ and $H$ ranks but omit writing them throughout the rest of paper.

\subsection{Softmax as a Cascade of Einsums}
\label{sec:einsums:softmax}

We now apply the same precise notation to the softmax.
A softmax~\cite{softmax} over a 1-tensor is traditionally expressed with the following equation:
\begin{align}
A_{m} = \frac{e^{I_m}}{\sum_k e^{I_k}}
\end{align}
In the context of attention, this operation becomes two dimensional and can be expressed using the following cascade with input $QK_{m,p}$:
\begin{align}
SN_{m,p} &= e^{QK_{m,p}} \label{eq:softmax:vanilla:sn}\\
SD_p &= SN_{m,p} \label{eq:softmax:vanilla:sd}\\
A_{m, p} &= SN_{m,p} / SD_p \label{eq:softmax:vanilla:a}
\end{align} 
For each point in the iteration space ($m$, $p$), we exponentiate $QK_{m,p}$ to generate the softmax numerator ($SN_{m,p}$ in Einsum~\ref{eq:softmax:vanilla:sn}), reduce $SN_{m,p}$ with addition to produce the softmax denominator ($SD_p$ in Einsum~\ref{eq:softmax:vanilla:sd}), and finally, divide the numerator and denominator to produce the final result ($A_{m, p}$ in Einsum~\ref{eq:softmax:vanilla:a}).

\subsubsection{Improving Numerical Stability}
\label{sec:numerical_stability}

Because $e^{QK_{m,p}}$ can easily become extremely large, the above formulation suffers from overflow.
Therefore, practical implementations~\cite{tensorflow, pytorch} often prefer the numerically stable variant that replaces Einsum~\ref{eq:softmax:vanilla:sn} with:
\begin{align}
GM_p &= QK_{m,p} :: \bigvee_{m} \text{max}(\cup) \label{eq:softmax:stable:m} \\
SN_{m,p} &= e^{QK_{m,p} - GM_p} \label{eq:softmax:stable:sn}
\end{align}
and drop the $\frac{1}{\sqrt{E}}$ term when computing $QK_{m,p}$.\footnote{The $\frac{1}{\sqrt{E}}$ term was introduced to bound the magnitude of $SN_{m, p}$~\cite{attention}. Because the numerically stable softmax variant already accomplishes this, the scaling is often omitted~\cite{flashattention, flashattention2, choi-et-al-2022}.}
To compute the \emph{global maximum}\footnote{``Global'' here refers to over the entire $M$ fiber.} $GM_p$, we reduce $QK_{m,p}$ with the operator $\text{max}$ (instead of $+$). Notice that subtracting $GM_p$ from $QK_{m,p}$ in the exponent is equivalent to dividing by $e^{GM_p}$, 
and because the $\frac{1}{e^{GM_p}}$ term appears in both the numerator ($SN_{m,p}$ via Einsum~\ref{eq:softmax:stable:sn}) and denominator ($SD_p$ via Einsum~\ref{eq:softmax:vanilla:sd}), the result ($A_{m, p}$) stays the same.
This construction improves numerical stability by bounding the values of the softmax numerator $SN_{m,p}$ to the range $(0, 1]$.

\subsection{Optimizing Softmax Compute}
\label{sec:einsums:opt-softmax-compute}

We now describe an optimization to attention that reduces compute requirements, specifically division.
This optimization was used in FlashAttention-2~\cite{flashattention2}.
We point out that it can be applied more broadly, i.e., to any cascade we discuss in Section~\ref{sec:einsums:opt-softmax-traffic}.
Einsum~\ref{eq:softmax:vanilla:a} requires $M \times P$ divisions.
While this is the best we can do for an independent softmax, we note that attention does not use the softmax in isolation~\cite{attention}. 
Instead, it subsequently multiplies the result, $A_{m, p}$, and another tensor, $V_{f, m}$, per Einsum~\ref{eq:attn:basic:av}, reproduced here:
\begin{align*}
AV_{f, p} = A_{m, p} \times V_{f, m}
\end{align*}
To optimize the full attention cascade, we can refactor Einsums~\ref{eq:softmax:vanilla:a} and~\ref{eq:attn:basic:av} by, instead, first combining $SN_{m,p}$ and $V_{f, m}$ 
(Einsum~\ref{eq:softmax:div:snv}) and reducing across the $M$ rank and then performing the division (Einsum~\ref{eq:softmax:div:av}), as follows:
\begin{align}
SNV_{f, p} &= SN_{m,p} \times V_{f, m} \label{eq:softmax:div:snv} \\
AV_{f, p} &= SNV_{f, p} / SD_p \label{eq:softmax:div:av} 
\end{align}
This reassociation does $F \times P$ divisions instead of $M \times P$ divisions.
Since $M$ is the sequence length and $F$ is an embedding dimension (i.e., $M \gg F$), this reassociation 
\emph{reduces} the required divisions (by a factor of $\frac{M}{F}$).

\subsection{Optimizing Softmax Live Footprint and Memory Traffic}
\label{sec:einsums:opt-softmax-traffic}

\begin{table}[t]
\centering
\begin{tabular}{|c|c|c|}
\hline
\textbf{3-pass} & \textbf{2-pass} & \textbf{1-pass} \\ \hline
PyTorch~\cite{pytorch} & TileFlow~\cite{tileflow} & FlashAttention~\cite{flashattention} \\
TensorFlow~\cite{tensorflow} & Choi et al.~\cite{choi-et-al-2022} & FlashAttention-2~\cite{flashattention2} \\ 
FLAT~\cite{flat} & & Rabe and Staats~\cite{rabe-and-staats-2022}  \\
E.T.~\cite{et} && \\
\hline
\end{tabular}
\caption{Classifying prior attention algorithms.
}
\label{tab:passes}
\vspace{-0.2in}
\end{table}

We now apply the analysis described in Section~\ref{sec:passes} to analyze attention's live footprint and memory traffic.
We consider the \emph{exact attention} literature, omitting works that either do not model/evaluate the softmax or include approximation strategies that improve performance at the cost of reduced accuracy (increased perplexity).
We discuss the latter in Section~\ref{sec:related}.

We find that existing approaches to attention can be classified as either 3-pass, 2-pass, or 1-pass cascades, where an $N$-pass cascade performs $N$ passes of a given $M$ fiber.
See Table~\ref{tab:passes}.
Next, we describe the key ideas of each.

\subsubsection{3-Pass Attention Cascades}

The 3-pass cascade is the straightforward, numerically stable cascade that we already discussed in Section~\ref{sec:numerical_stability}, namely Einsums~\ref{eq:softmax:stable:m}-\ref{eq:softmax:stable:sn} followed by Einsums~\ref{eq:softmax:vanilla:sd}-\ref{eq:softmax:vanilla:a}, reproduced in Cascade~\ref{fig:3pass} for clarity.

\begin{cascade}[H] 
    \caption{\label{fig:3pass} The 3-pass attention cascade.}
    \begin{mdframed}
    \begin{align}
QK_{m, p} &= Q_{e, p} \times K_{e, m} &\text{\cascadecomment{ Pass 1 }} \label{eq:3pass:qk}\\
GM_p &= QK_{m,p} :: \bigvee_{m} \text{max}(\cup) \label{eq:3pass:gm}\\
SN_{m,p} &= e^{QK_{m,p} - GM_p} &\text{\cascadecomment{ Pass 2 }} \label{eq:3pass:sn}\\
SD_p &= SN_{m,p} \label{eq:3pass:sd}\\
A_{m, p} &= SN_{m,p} / SD_p &\text{\cascadecomment{ Pass 3 }} \label{eq:3pass:a} \\
AV_{f, p} &= A_{m, p} \times V_{f, m} \label{eq:3pass:av}
    \end{align}
    \end{mdframed}
\end{cascade}

In Pass 1, we compute $QK_{m, p}$ and $GM_p$; in Pass 2, we compute $SN_{m, p}$ and $SD_p$; and in Pass 3, we compute $A_{m, p}$ and $AV_{f, p}$.
Notice that we must finish an entire $M$ fiber of Einsum~\ref{eq:3pass:gm} (reading an entire $M$ fiber of $QK_{m, p}$)
before $GM_p$ is ready to start Einsum~\ref{eq:3pass:sn} (where we must read the same $M$ fiber of $QK_{m, p}$ again).
Similarly, we must finish an entire $M$ fiber of Einsum~\ref{eq:3pass:sd} (reading an entire $M$ fiber of $SN_{m, p}$)
before $SD_p$ is ready to start Einsum~\ref{eq:3pass:a} (where we must read the same $M$ fiber of $SN_{m, p}$ again).
To summarize, as a consequence of the \emph{dependencies} between Einsums, this cascade must perform three passes over each $M$ fiber.
This holds for any choice of mapping (including ones that perform fusion).

\subsubsection{2-Pass Attention Cascades}
\label{sec:einsums:2pass}

We now briefly summarize the 2-pass cascade, deferring details due to space.
Rather than computing the global max and then starting the softmax (as in the 3-pass cascade), the 2-pass cascade first partitions the input, computes a per-partition \emph{local max} and applies it to form a variant of $SN_{m,p}$ whose elements are likewise partitioned and adjusted by the local max.
Analogously, each partition gets a local denominator (also adjusted by the same local max).
While this is occurring, it builds the global max from the local max values.
Next, in a second pass, it uses the global max to correct the per-partition numerators and denominators and compute the softmax output.

\subsubsection{1-Pass Attention Cascades}
\label{sec:einsums:1pass}
While prior work proposes multiple different 1-pass 
cascades~\cite{flashattention, flashattention2, rabe-and-staats-2022}, the main ideas are the same in each. 
Rather than using the per-partition local max to compute the local numerator and denominator, instead keep a \emph{running max} that represents the max value seen so far. 
Each time a new running max is computed, also adjust previous results (e.g., numerator-times-$V$, denominator, etc.) with this max.

This transformation can be described more precisely using the reassociations presented in Section~\ref{sec:passes:reassociation}.
First, we modify 
Cascade~\ref{fig:3pass} to multiply the softmax numerator-times-$V$ and then compute the division (as described in Section~\ref{sec:einsums:opt-softmax-compute}). 
This reassociation combines the second and third passes of Cascade~\ref{fig:3pass} (see Section~\ref{sec:passes:reassociation:defer}).
To ensure numerical stability, we cannot use the same strategy to combine the first and second passes. 
So we instead use the iterative approach (see Section~\ref{sec:passes:reassociation:iter}).

We are now ready to describe FlashAttention-2's 1-pass cascade (shown as Cascade~\ref{fig:1pass}). 
We later use it to build \name.
Note the evidently increased compute relative to the 3-pass cascade.
We will carefully design the binding in Section~\ref{sec:mapping} to hide these overheads on a spatial architecture.

\begin{cascade}[h] 
    \caption{\label{fig:1pass} A 1-pass attention cascade.
    Note that $M1$ is used as a standard rank (e.g., to access $BQK_{m1, m0, p}$) and as an iterative rank (e.g., to access $RM_{m1, p}$).
    The stopping condition for all iterative ranks is $m1 \geq M1$ (Statement~\ref{eq:stop}).
    }
    \begin{mdframed}
    
Initialization:
\begin{align}
BK_{e, m1, m0} &= K_{e, m1 \times M0 + m0} \label{eq:softmax:iter:bk} \\
BV_{f, m1, m0} &= V_{f, m1 \times M0 + m0} \label{eq:softmax:iter:bv} \\
RM_{m1:m1=0, p} &= -\infty \label{eq:softmax:iter:rm:0} \\
RD_{m1:m1=0, p} &= 0 \label{eq:softmax:iter:rd:0} \\
RNV_{m1:m1=0, p} &= 0 \label{eq:softmax:iter:rnv:0}
\end{align}

Extended Einsums:
    \begin{align}
BQK_{m1, m0, p} &= Q_{e, p} \times BK_{e, m1, m0} \label{eq:softmax:iter:qk} \\
LM_{m1, p} &= BQK_{m1, m0, p} :: \bigvee_{m0} \text{max}(\cup) \label{eq:softmax:iter:lm} \\
RM_{m1 + 1, p} &= max(RM_{m1, p}, LM_{m1, p}) \label{eq:softmax:iter:rm} \\
SLN_{m1, m0, p} &= e^{BQK_{m1, m0, p} - RM_{m1 +1, p}} \label{eq:softmax:iter:sln}\\
SLD_{m1, p} &= SLN_{m1, m0, p} \label{eq:softmax:iter:sld}\\
SLNV_{f, m1, p} &= SLN_{m1, m0, p} \times BV_{f, m1, m0} \label{eq:softmax:iter:slnv} \\
PRM_{m1, p} &= e^{RM_{m1, p} - RM_{m1 + 1, p}} \label{eq:softmax:iter:prm}\\
SPD_{m1, p} &= RD_{m1, p} \times PRM_{m1, p} \label{eq:softmax:iter:spd} \\
RD_{m1 + 1, p} &= SLD_{m1, p} + SPD_{m1, p} \label{eq:softmax:iter:rd} \\
SPNV_{f, m1, p} &= RNV_{f, m1, p} \times PRM_{m1, p} \label{eq:softmax:iter:spnv} \\
RNV_{f, m1 + 1, p} &= SLNV_{f, m1, p} + SPNV_{f, m1, p} \label{eq:softmax:iter:rnv} \\
AV_{f, p} &= RNV_{f, M1, p} / RD_{M1, p} \label{eq:softmax:iter:av} \\
&\diamond: m1 \geq M1 \label{eq:stop}
    \end{align}
    \end{mdframed}
\end{cascade}

We will start by expressing the partitioning of both of the inputs $K_{e,m}$ and 
$V_{f, m}$ into M1 chunks of M0 elements each (Einsums~\ref{eq:softmax:iter:bk}-\ref{eq:softmax:iter:bv}).
After computing $BQK_{m1, m0, p}$, this allows us to perform operations like maximum on individual $M0$ fibers, rather than on the whole tensor (Einsum~\ref{eq:softmax:iter:lm}).
The problem is, of course, that the local maximum is not necessarily the same for all $M0$ fibers and so will not just cancel nicely like the global maximum.

We resolve this by instead using the running maximum ($RM_{m1, p}$)---the global maximum of all inputs seen so far---instead of the local maximum.
We recognize that $M1$ can also serve as an iterative rank, and iteratively build up $RM_{m1, p}$.
After initializing $RM_{m1:m1=0, p}$ to $-\infty$ (Einsum~\ref{eq:softmax:iter:rm:0}), we compute a new running maximum $RM_{m1 + 1, p}$ using the running maximum computed in the previous iteration $RM_{m1, p}$ and the new local maximum $LM_{m1, p}$ (Einsum~\ref{eq:softmax:iter:rm}).

We can now use the running maximum to compute a local numerator $SLN_{m1, m0, p}$ (Einsum~\ref{eq:softmax:iter:sln}), a local denominator $SLD_{m1, p}$ (Einsum~\ref{eq:softmax:iter:sld}), and even the softmax numerator-times-$V$ $SLNV_{f, m1, p}$ (Einsum~\ref{eq:softmax:iter:slnv}) using the partitioned $BV_{f, m1, m0}$ (Einsum~\ref{eq:softmax:iter:bv}). 

Now consider the softmax denominator. 
Eventually, we would like to reduce $SLD_{m1, p}$ into a 1-tensor, but because its values may have been computed with different maximums, we cannot simply use addition.
Instead, by introducing a new running denominator $RD_{m1, p}$ with iterative rank $M1$, we can correct the old denominator $RD_{m1, p}$ to the new running maximum $RM_{m1 + 1, p}$ and then perform the addition.
We start by initializing the running denominator at the start of the computation to 0 (Einsum~\ref{eq:softmax:iter:rd:0}).
Then, at each point $m1$, the correction factor $PRM_{m1, p}$ allows us to correct the previous running denominator $RD_{m1, p}$ with the new maximum (Einsum~\ref{eq:softmax:iter:spd}).
In other words, $RD_{m1, p}$ is downscaled by $e^{RM_{m1, p}}$.
$SPD_{m1, p}$ ``switches’' the downscaling factor on $RD_{m1, p}$ to $e^{RM_{m1 + 1, p}}$ by multiplying $RD_{m1, p}$ by $\frac{e^{RM_{m1, p}}}{e^{RM_{m1 + 1, p}}}$ ($PRM_{m1, p}$).
Once $SLD_{m1, p}$ and $SPD_{m1, p}$ have the same maximum, they can be combined to produce the new running denominator $RD_{m1 + 1, p}$ (Einsum~\ref{eq:softmax:iter:rd}).
We can do the same to compute the running numerator-times-$V$ (Einsums~\ref{eq:softmax:iter:rnv:0},~\ref{eq:softmax:iter:spnv}-\ref{eq:softmax:iter:rnv}).

Finally, $AV_{f, p}$ can be computed by dividing the final numerator-times-$V$ by the final denominator.
By construction, at this point, $RNV_{f, M1, p}$ and $RD_{M1, p}$ are both downscaled by the same maximum $RM_{M1, p}$ (conveniently, also the global maximum) and can be correctly combined.

\section{Mapping and Binding Attention}
\label{sec:mapping}

Based on the framework from Section~\ref{sec:einsums}, we now describe \name, an efficient mapping and binding of an attention algorithm (specifically the 1-pass cascade in Cascade~\ref{fig:1pass}) to a spatial array-style architecture.
To enable maximum flexibility while binding, \name's mapping places each iteration space point in its own logical task.

The goal when binding a cascade onto hardware is to fully utilize all available compute units.
In our evaluation of prior work (Figure~\ref{fig:eval:util} and Section~\ref{sec:eval:attn}), we observe that at short sequence lengths, the 2D PE array is under-utilized because it must wait for the 1D PE array to compute the softmax.
At longer sequence lengths, both arrays are under-utilized since the workload becomes memory-bandwidth limited.

\name's binding addresses these issues to achieve full utilization on both the 1D and 2D PE arrays.
First, we decrease the compute performed by the 1D array by (1) applying the division reduction optimization (Section~\ref{sec:einsums:opt-softmax-compute}) and (2) sharing the other operations (sum/max/exp) between the 1D and 2D arrays.
Similarly, we ensure that the workload is never memory-bandwidth limited by deeply fusing all Einsums in the cascade to restrict the live footprint to only what can be buffered on-chip.
No matter the sequence length, our dataflow is never forced to spill any of its intermediates off-chip.

\begin{figure}[h]
    \centering
    \includegraphics[width=.35\textwidth]{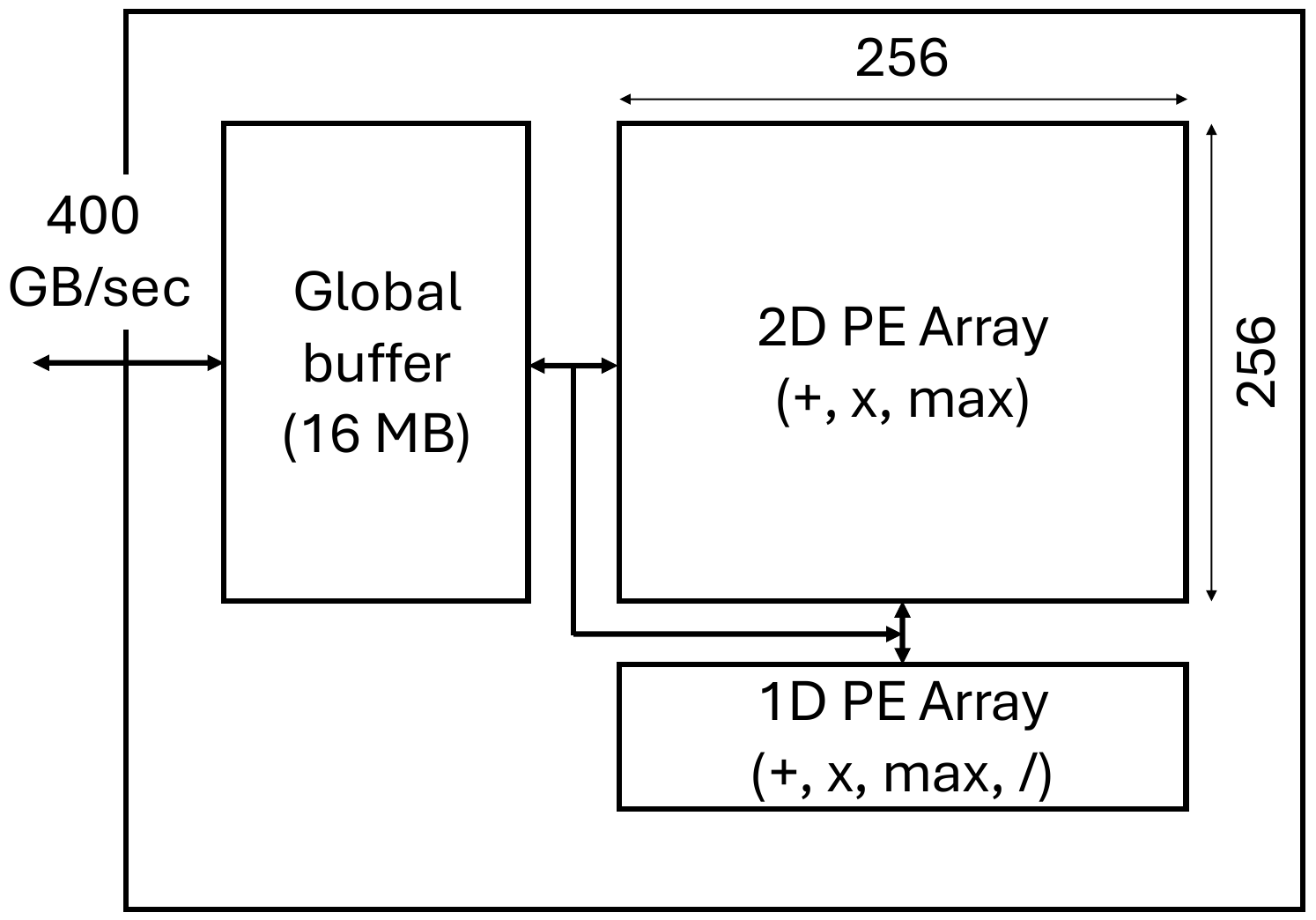}
    \caption{
    Spatial array architecture assumed for \name.
    }
    \label{fig:proposal:arch}
\end{figure}

\textbf{Architecture.}
\name is a spatial array architecture based on the TPUv2/TPUv3~\cite[Figure 1(e)]{tpu-v2-v3}.
The off-chip DRAM and a large global buffer both feed data to connected 2D and 1D arrays (see Figure~\ref{fig:proposal:arch}).
We set parameters to match the cloud configuration in prior work~\cite{flat}.

\begin{figure}[t!]
    \centering
    \begin{subfigure}[t]{0.15\textwidth}
        \centering
        \includegraphics[width=\textwidth]{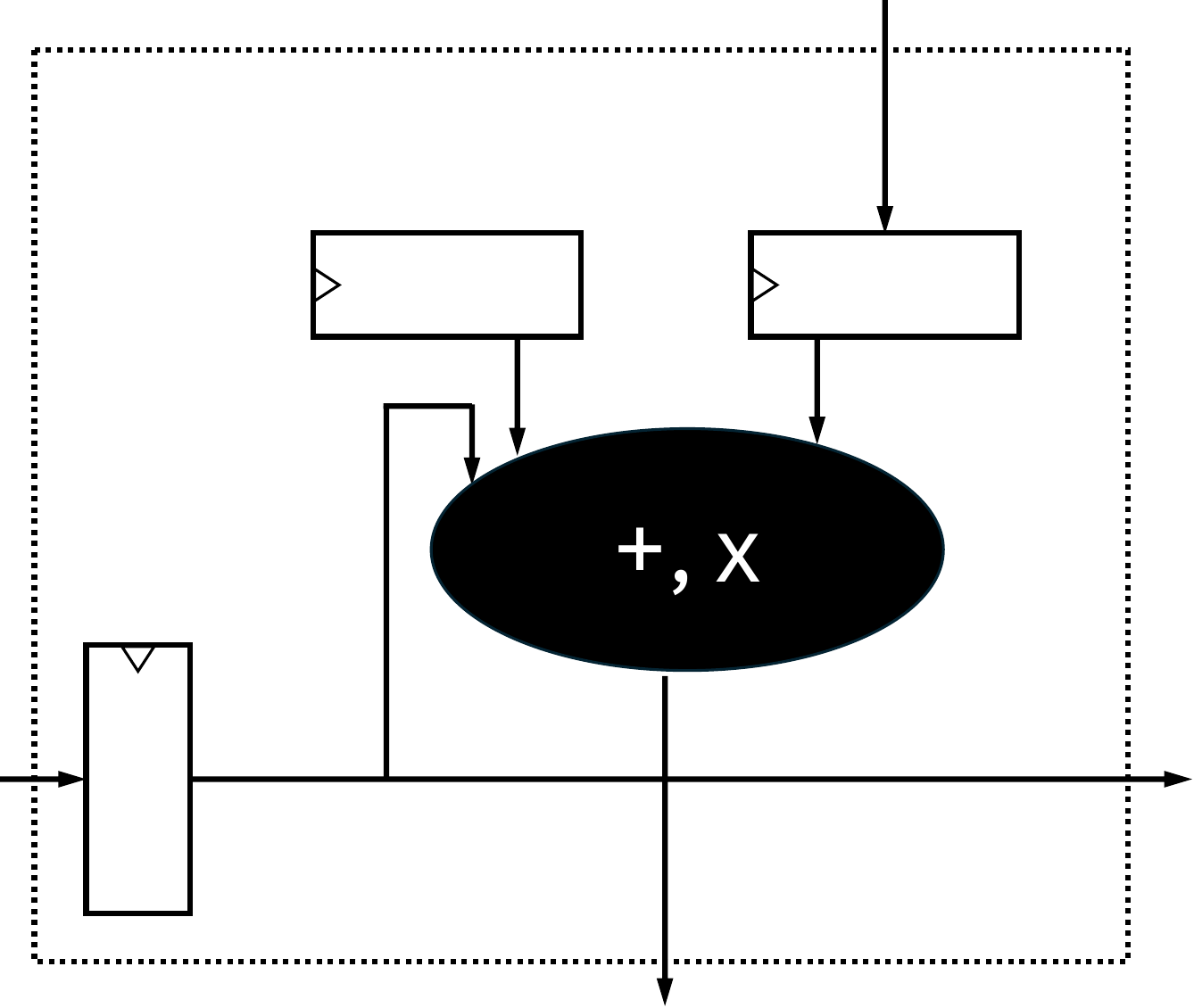}
        \caption{TPU~\cite{tpu} PE}
        \label{fig:pe:tpu}
    \end{subfigure}
    ~
    \begin{subfigure}[t]{0.15\textwidth}
        \centering
        \includegraphics[width=\textwidth]{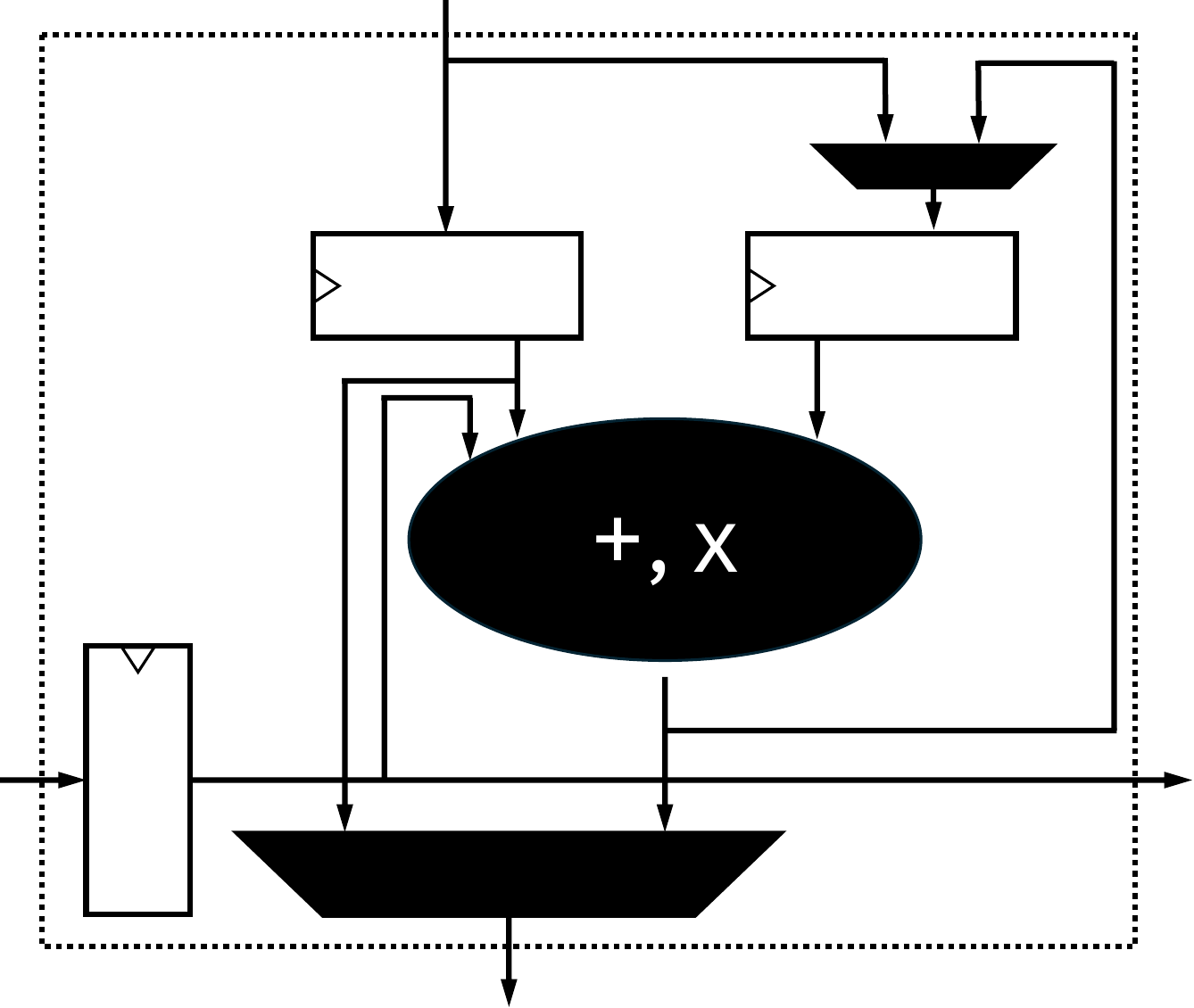}
        \caption{FLAT~\cite{flat} PE}
        \label{fig:pe:flat}
    \end{subfigure}
    ~
    \begin{subfigure}[t]{0.15\textwidth}
        \centering
        \includegraphics[width=\textwidth]{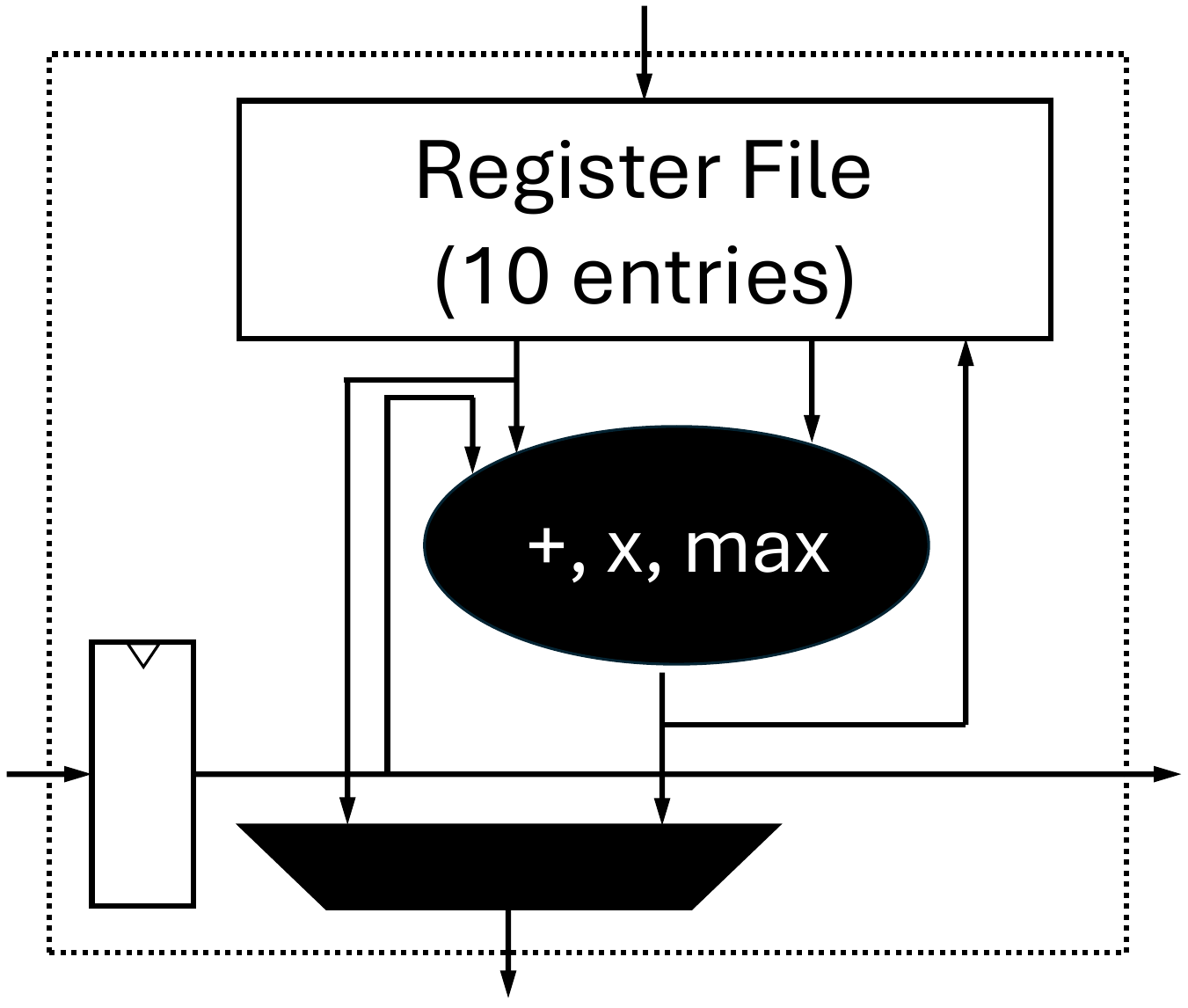}
        \caption{\name PE}
        \label{fig:pe:fusemax}
    \end{subfigure}
    ~
    \caption{2D PE architecture evolution.}
    \label{fig:pe}
\end{figure}

Figure~\ref{fig:pe} shows the evolution of the 2D PE array architecture, from a fixed-dataflow multiply-accumulate TPU PE (Figure~\ref{fig:pe:tpu}) to a flexible-dataflow multiply-accumulate PE (Figure~\ref{fig:pe:flat}) to a \name PE (Figure~\ref{fig:pe:fusemax}).
Note, although both the 1D and 2D PE arrays in \name perform exponentiation, we implement exponentiation with 6 sequential multiply-accumulate operations~\cite{Nilsson:2014:HIE, spatten} and therefore do not require a dedicated exponentiation unit.

\textbf{Mapping.}
Prior attention accelerators~\cite{flat, tileflow} explore fusing many of attention's loop nests together.
However, because these accelerators all use multi-pass cascades, the algorithmic minimum live footprint of some tensors (e.g., $QK_{m, p}$) is $O(M)$, meaning that for long sequence lengths, intermediates cannot be buffered on chip.

\name leverages fusion in conjunction with the 1-pass cascade to eliminate the memory traffic of these tensors, regardless of the sequence length.
Specifically, we partition on both $M$ and $P$ (forming $M1,M0$ and $P2,P1,P0$), and maximally fuse all levels in the attention loopnest as shown in Mapping~\ref{fig:loopnest}.
That is, all Einsums in Cascade~\ref{fig:1pass} are fused except for the last (which is fused to the rest only on $P2$).

\begin{mapping} 
    \caption{\label{fig:loopnest}
    The \name mapping as a loopnest.
    We partition on both $M$ and $P$ and map the innermost ranks $M0$ and $P0$ to the spatial array PEs.
    \texttt{ComputeRNVTile} performs Einsums~\ref{eq:softmax:iter:qk}-\ref{eq:softmax:iter:rnv} from Cascade~\ref{fig:1pass}.
    \texttt{ComputeAVTile} performs Einsum~\ref{eq:softmax:iter:av}. 
    Note that each Einsum represents a loopnest: by writing all Einsums in \texttt{ComputeRNVTile} under a single loopnest, we mean that we are maximally fusing those loopnests.
    Outer loops over $B$ and $H$ (if performing batched multihead attention) are not shown.
    }
    \begin{mdframed}
{\small
    \begin{verbatim}
for p2 ...:
  for m1 ...:
    for p1 ...:
      parallel_for p0 ...:
        parallel_for m0 ...:
          (RNV[:, m1 + 1, p2, p1, p0], 
           RD[m1 + 1, p2, p1, p0]) = 
              ComputeRNVTile(
                Q[:, p2, p1, p0], 
                K[:, m1, m0], V[:, m1, m0])
  for p1 ...:
    parallel_for p0 ...:
      AV[:, p2, p1, p0] = 
        ComputeAVTile(
          RNV[:, m1 + 1, p2, p1, p0], 
          RD[m1 + 1, p2, p1, p0])
    \end{verbatim}
}
    \end{mdframed}
\end{mapping}

While prior work implementing attention in hardware~\cite{flat, tileflow} does utilize the 2D spatial array for the tensor products, it fails to do so for the softmax, choosing instead to use the 1D array.
Because there are far fewer total PEs in the 1D array than the 2D array, the softmax becomes a bottleneck.
\name improves utilization of the 2D spatial array by using it for both the tensor products and the exponentiation operator in the softmax.
\name parallelizes across the $M0$ and $P0$ ranks throughout the attention kernel (see Mapping~\ref{fig:loopnest}).
We set $M0\times P0=\#\;\mathrm{2D\;Array}\;\mathrm{PEs}$.
The large spatial reductions required when parallelizing across the $M0$ rank are easily handled by the low-cost inter-PE communication network.

\begin{figure*}[h]
    \centering
    \includegraphics[width=.8\textwidth]{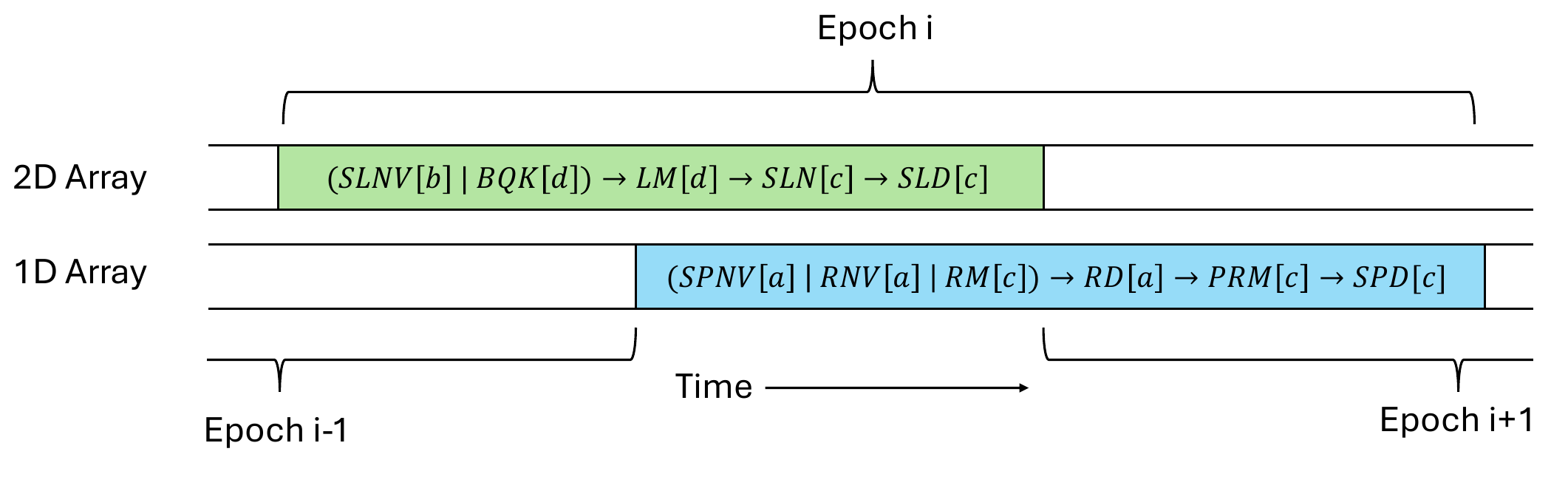}
    \caption{
    \name pipelining at a glance.
    Each tensor name (e.g., $SLNV$) corresponds to the Einsum used to compute that tensor (see Cascade~\ref{fig:1pass}).
    $a$, $b$, $c$ and $d$ denote tile-relative coordinates where $a<b<c<d$.
    If Epoch $i$ produces tiles with coordinates $a,b,c,d$, Epoch $i+1$ produces tiles with identifiers $a+1,b+1,c+1,d+1$.
    And so on.
    `$A|B$' denotes `computing tile $A$ is interleaved with computing tile $B$.'
    `$A\rightarrow B$' denotes `computing tile $A$ is done before computing tile $B$.'
    Computing $AV_{f,p}$ is not shown.
    The green and blue time periods making up an epoch take almost the same number of cycles.
    }
    \label{fig:proposal:waterfall}
\end{figure*}

\textbf{Binding.}
The dependencies between different Einsums in our cascade necessitate a binding that implements fine-grain pipeline parallelism to achieve high utilization of both the 1D and 2D spatial arrays.
Figure~\ref{fig:proposal:waterfall} shows the waterfall diagram for \name in the steady state.
Time is broken into epochs.
Each epoch performs the same set of tile-granular operations at specific tile-relative coordinates (given by $a,b,c,d$ in the figure). 
Across all epochs, the kernel evaluates all tiles and each Einsum in Cascade~\ref{fig:1pass} is mapped to either the 2D or 1D array for all epochs (as shown in the figure).

A major design consideration when binding the attention kernel is how to overcome the latency of fills and drains to/from the spatial array.
Consider a tile of $QK_{m,p}$ of shape $M0\times P0$.
Per Einsum~\ref{eq:attn:basic:qk}, the iteration space to evaluate this tile is $E\times M0 \times P0$ which becomes $E$ cycles on the spatial array.
For the networks we evaluate, $E=64$ or $128$.
Assume $E=64$.
Using an output stationary dataflow, while each PE performs 64 MACCs, it takes $\sim256$ cycles to both fill and drain the spatial array.  
Without careful interleaving, this combination of parameters causes low utilization because,
for example, the running max $RM_{m1 + 1, p1, :}$ cannot be computed until a tile of $QK_{m1, :, p1, :}$ is completed and spatially reduced (drained) to form the local max $LM_{m1, p1, :}$ (Einsums~\ref{eq:softmax:iter:lm}-\ref{eq:softmax:iter:rm}).

Our binding address the above issues with two levels of interleaving.
First, we interleave the construction of dependent tiles across epochs. 
This is reminiscent of software pipelining.
For example, in Figure~\ref{fig:proposal:waterfall} the $d$-th tile of $BQK$ and $LM$ are completed in Epoch $i$ (as they correspond to a fill followed by a drain and can be easily pipelined).
The $RM$ (which has to wait for the drain) for tile $d$ takes place \emph{in a later epoch}. 
Instead, Epoch $i$ computes an earlier tile's running maximum $RM[c]$.

\begin{figure*}[h]
    \centering
    \includegraphics[width=0.8\textwidth]{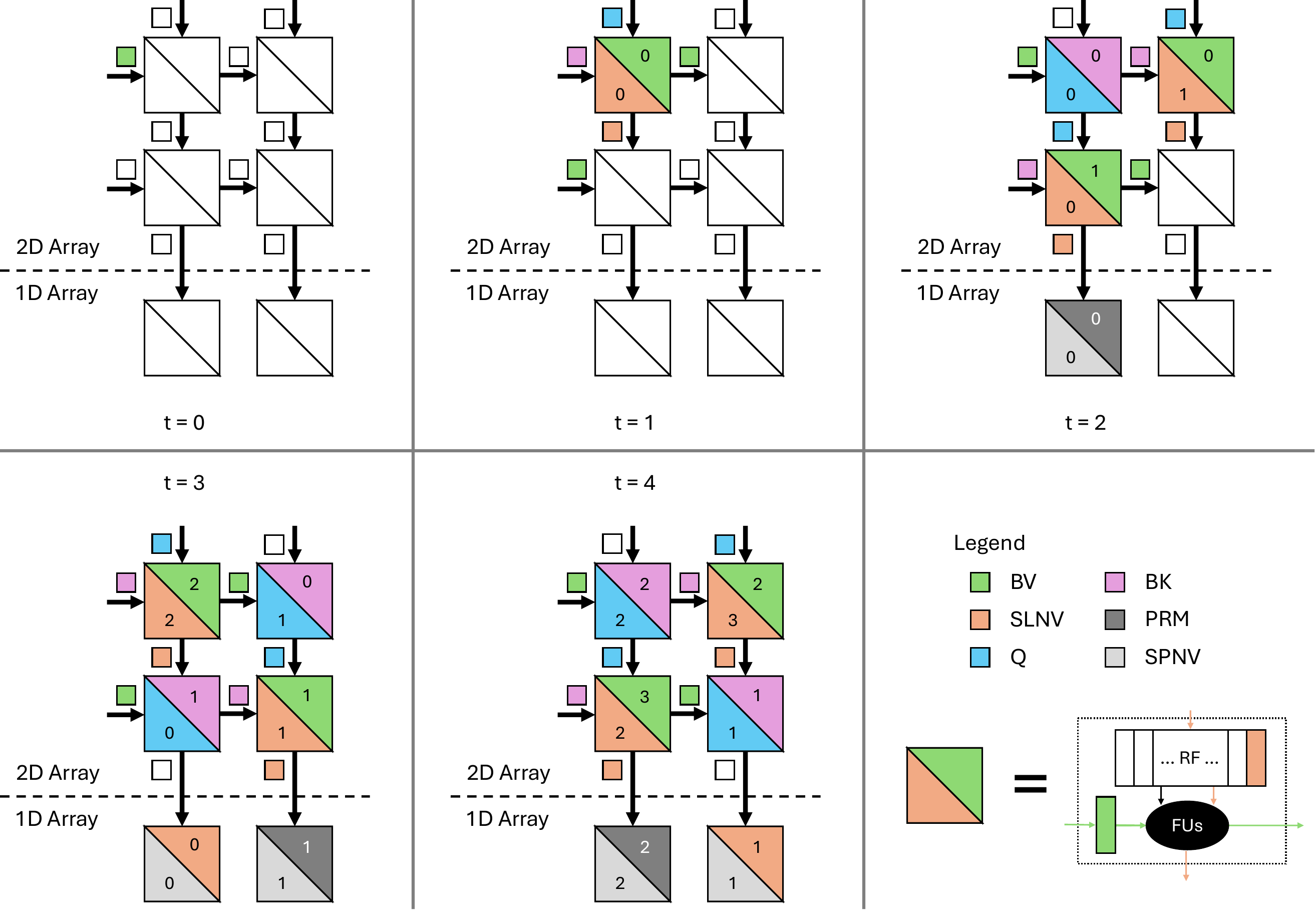}
    \caption{
    Intial pipeline fill ($t=0$ to $t=2$) and steady-state ($t=3$ and $t=4$) for the intra-epoch interleaving of $SLNV|BQK$ and $SPNV|RNV$ to maximize 2D and 1D PE utilization, respectively, on a toy 2x2 array.
    Each color indicates a tensor and each number indicates a point in that tensor (e.g., the point $BV_0$ moves from the top left PE at $t = 1$ to the top right PE at $t = 2$).
    To reason about signal timing, we use input (but not output) latches for data in each PE, so moving data appears on output wires.
    Some stationary tensors (e.g., $BQK$) and Einsums (e.g., $SLD$) are omitted for clarity.
    }
    \label{fig:proposal:interleaving}
\end{figure*}

Second, we interleave the construction of certain tiles within an epoch at a fine (e.g., cycle-by-cycle) granularity.
See the notation `$A|B$' in Figure~\ref{fig:proposal:waterfall}.
This is to ensure high utilization of both the 2D and 1D PE arrays at all times.
To make this more clear, Figure~\ref{fig:proposal:interleaving} shows the start up and steady-state interleaving of $SLNV$ and $BQK$ in the 2D array and $SPNV$ and $RNV$ in the 1D array.
In each cycle, a given PE in the 2D array computes a value for either $BQK$ or $SLNV$ and this alternates cycle by cycle.
Each neighbor-neighbor link in the array is active in every cycle---carrying data for one of the two operation types.
By interleaving $SLNV$ with $BQK$, the 1D PEs can concurrently compute $SPNV$ and $RNV$.

Putting everything together, as Section~\ref{sec:eval:attn} will show, the above enables high 
utilization of all 2D and 1D array PEs.

\textbf{\name on GPUs.}
\name's mapping and binding cannot be directly applied to GPUs.
\name's architecture features heterogeneous PEs, each with smaller per-PE storage, and cheap (but restricted) inter-PE communication.
Specifically, the networks that connect the PEs within the 2D array allow efficient, fixed-latency communication primarily between neighbors, including between the bottom of the 2D array and the 1D array.
However, the GPU architecture features opposite characteristics: homogeneous PEs, each with relatively large per-PE storage, and expensive, loosely coupled inter-PE communication.
While concurrent work~\cite{flashattention3} has explored using the GPU's Tensor Cores to compute $BQK$ and $SLNV$ and using software pipelining to hide the latency of the other compute, the GPU's loosely coupled threads require frequent synchronization to maintain correctness.
\name takes advantage of the tight coupling between the 2D and 1D arrays to statically schedule compute between the arrays, enabling high utilization across the board without sychronization.
\section{Evaluation}
\label{sec:eval}

\begin{figure*}[t!]
    \centering
    \begin{subfigure}[t]{\textwidth}
        \centering
        \includegraphics[width=\textwidth]{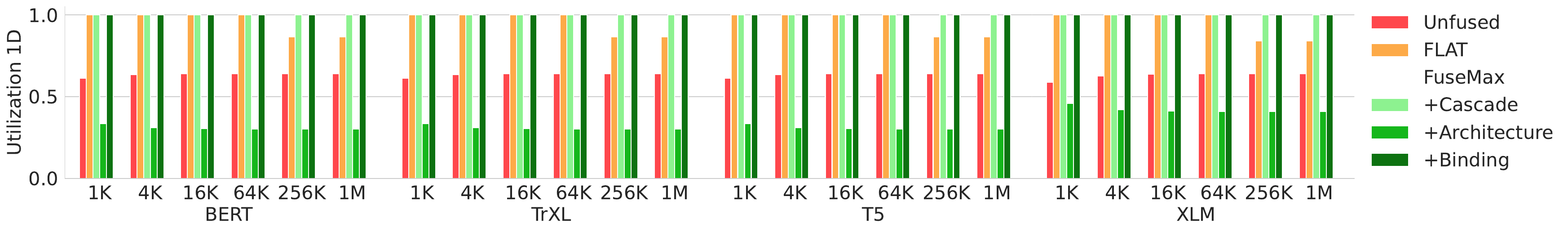}
        \caption{1D PE array utilization}
        \label{fig:eval:util:1d}
    \end{subfigure}
    
    \begin{subfigure}[t]{\textwidth}
        \centering
        \includegraphics[width=\textwidth]{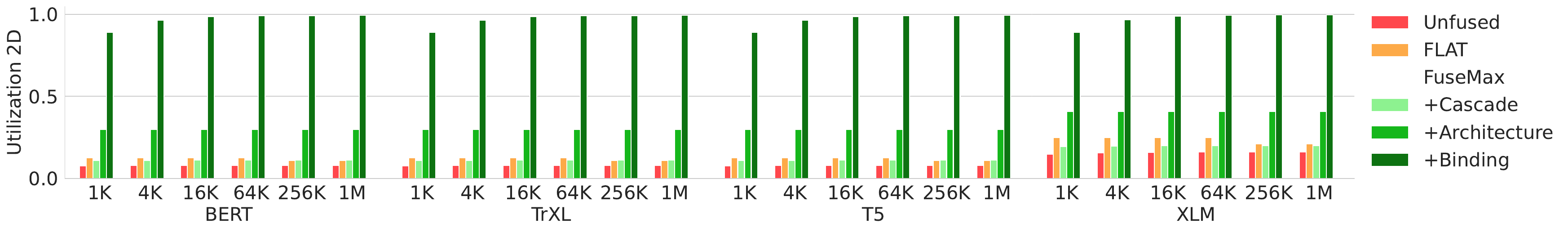}
        \caption{2D PE array utilization}
        \label{fig:eval:util:2d}
    \end{subfigure}%
    \caption{Utilization of the different PE arrays on the unfused baseline, FLAT, and three configurations building up \name.}
    \label{fig:eval:util}
\end{figure*}

\begin{figure*}[t!]
    \includegraphics[width=\textwidth]{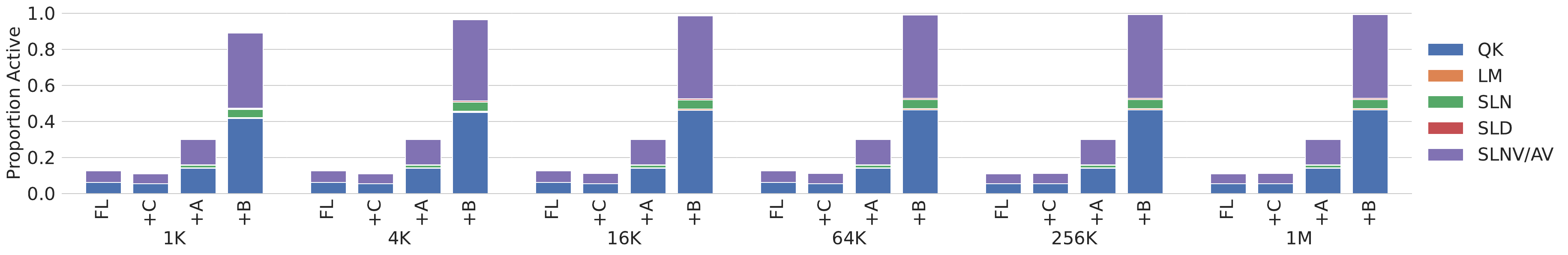}
    \caption{2D array utilization by Einsum across different configurations---FLAT (FL), +Cascade (+C), +Architecture (+A), and +Binding (+B)---and sequence lengths on BERT.}
    \label{fig:eval:util:breakdown}
\end{figure*}

\begin{figure*}[t!]
    \includegraphics[width=\textwidth]{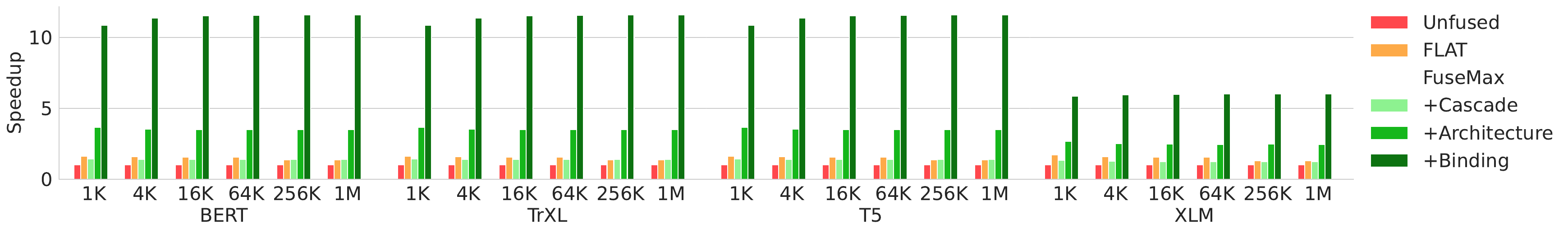}
    \caption{Speedup of attention for FLAT and three configurations building up \name over an unfused baseline.}
    \label{fig:eval:attn:speedup}
\end{figure*}

\begin{figure*}[t!]
    \includegraphics[width=\textwidth]{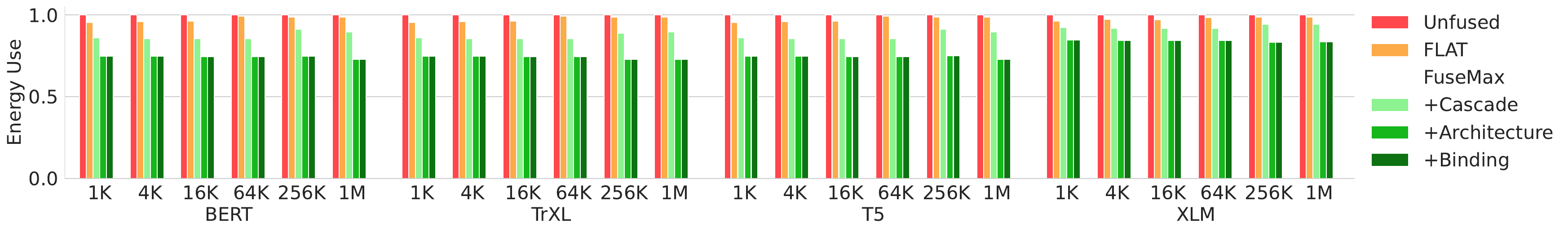}
    \caption{Energy consumption of attention for FLAT and three configurations building up \name over an unfused baseline.}
    \label{fig:eval:attn:energy}
\end{figure*}

\begin{figure*}[t!]
    \includegraphics[width=\textwidth]{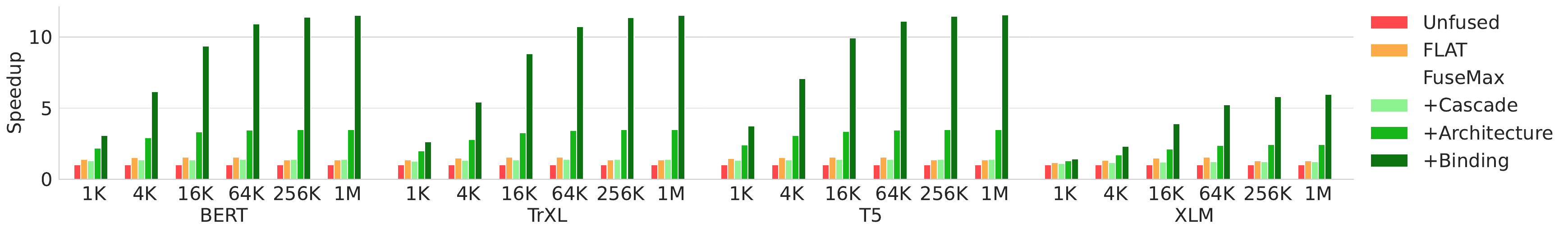}
    \caption{Speedup of transformer inference on FLAT and three configurations building up \name over an unfused baseline.}
    \label{fig:eval:end2end:speedup}
\end{figure*}

\begin{figure*}[t!]
    \includegraphics[width=\textwidth]{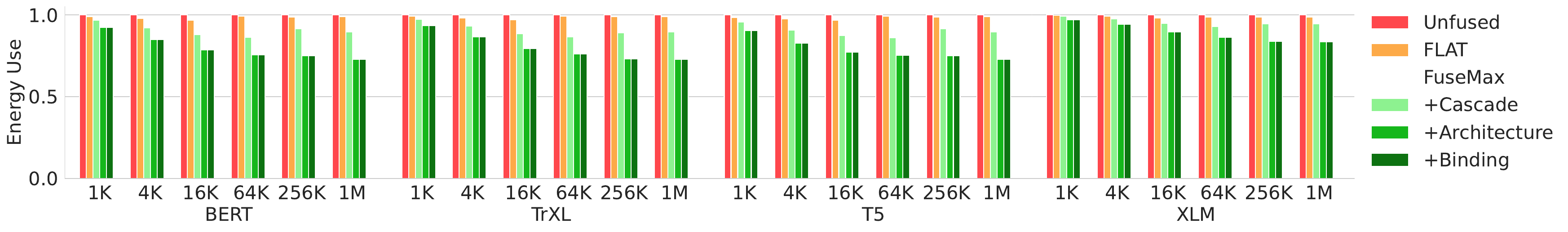}
    \caption{Energy consumption of transformer inference on FLAT and three configurations building up \name over an unfused baseline.
    }
    \label{fig:eval:end2end:energy}
\end{figure*}

In this section, we demonstrate how \name's cascade, architecture, and binding work together to achieve improvements in both performance and energy relative to the state of the art, for both attention and end-to-end transformer inference.

\subsection{Experimental Set-Up}
\label{sec:eval:setup}

First, we present the experimental setup details common to all following subsections.

\textbf{Workloads.}
We evaluate all accelerators and configurations using the same transformer models used by FLAT~\cite{flat}:
BERT-Base~\cite{bert} (BERT), TrXL-wt103~\cite{trxl-xlm} (TrXL), T5-small~\cite{t5} (T5), and XLM~\cite{trxl-xlm}.
We omit FlauBERT~\cite{flaubert} because it uses the same hyperparameters as TrXL.
We also note that though T5 is an encoder-decoder model, we only evaluate the encoder in this work.
Following FLAT, we use a batch size $B = 64$ for all evaluations.

\textbf{Modeling with Timeloop and Accelergy.}
We perform our evaluation using two tools for tensor algebra accelerator modeling and design space exploration: Timeloop~\cite{timeloop} and Accelergy~\cite{accelergy}.
We use these tools to build models of the accelerator architectures at a 45nm technology node and evaluate each Einsum individually.
Results from individual Einsums are combined using heuristics presented in prior work for evaluating full cascades~\cite{teaal}.
Together, these tools allow us to evaluate execution time, energy, and area for all our designs.
We perform floating-point division using the design in Xia et al.~\cite{Xia:2021:LLB}, scaled down to a 45nm technology node~\cite{accelergy}.

\textbf{Unfused Baseline.}
We build the unfused baseline by combining the costs of three phases: $QK$ (Einsum~\ref{eq:attn:basic:qk}), the 3-pass softmax (Cascade~\ref{fig:3pass}), and $AV$ (Einsum~\ref{eq:attn:basic:av}).
Because this baseline is unfused, each phase can be scheduled independently, but proceed sequentially and require outputs to be written to memory between phases.
We use Timeloop to search for efficient mappings to perform $QK$ and $AV$.
Additionally, we model the softmax for the unfused baseline by allowing the accelerator to load the $M$ fibers of the input on-chip one-by-one (spilling if there is not enough space) before performing the compute.
We model the memory traffic, compute, and energy required to perform all Einsums required for attention.

\textbf{FLAT Baseline.}
Our main baseline is the state-of-the-art attention accelerator FLAT~\cite{flat}.
Though we started with the FLAT authors' original code, we found and corrected a number of bugs.
Through private correspondence with the FLAT authors, we verified the bugs were indeed bugs.
We also discovered a couple of larger conceptual errors, which the authors told us to avoid by restricting FLAT to only search through configurations without these issues.

Beyond correcting the FLAT codebase, we created and validated a Timeloop model that reproduces the FLAT authors' (corrected) code to within $<1\%$ error.
However, the FLAT codebase does not model the cost to perform the softmax.
Specifically, their model ignores the cost of the data transfers required for the softmax (between any levels of the memory hierarchy) and uses $2^{30}$ 1D PEs for compute.
When comparing \name and FLAT in this work, we augment our Timeloop model to model softmax correctly per the 3-pass cascade implicitly assumed by FLAT using only 256 1D PEs.

\textbf{\name Configurations.}
To demonstrate the sources of the improvements achieved by \name, we present three configurations, one associated with each of the major changes we propose:
+Cascade uses the 1-pass cascade on the FLAT architecture, +Architecture adds the \name architecture but implements a binding that fully produces and consumes one $M0 \times P0$ tile of $BQK$ before starting the next, and +Binding adds \name's pipelined/interleaved binding.

\textbf{Hardware parameters.}
Figure~\ref{fig:proposal:arch} shows the selected hardware parameters.
We chose the PE array dimension to match FLAT's cloud accelerator and then set the global buffer capacity so that the overall chip area was as close to FLAT's as possible.
Also following FLAT, we use a 940 MHz frequency.
We use Accelergy to model the area of both designs and find that \name is 6.4\% smaller.

\subsection{Evaluating Attention}
\label{sec:eval:attn}

We now evaluate \name to demonstrate the benefits it provides on the attention kernel by comparing it to the two baselines.

\textbf{Utilization.}
Figure~\ref{fig:eval:util:1d} shows the utilization of the 1D PE array when performing attention.
FLAT's utilization drops for sequence lengths $\geq 256\text{K}$---it becomes memory bandwidth limited because it must spill the $QK$ and $A$ tensors to memory.
By using a 1-pass cascade (+Cascade), \name's utilization becomes independent of sequence length.
We also note that without the \name binding (+Architecture), the 1D array is forced to stall and utilization drops.
Adding in this binding (+Binding) enables \name to fully utilize the 1D array again.

Similarly, Figure~\ref{fig:eval:util:2d} shows the utilization of the 2D array.
Because of the large amount of compute required for the softmax, most configurations achieve poor utilization of this array.
In fact, because the 1-pass cascade increases the compute required, +Cascade's 2D array utilization is lower than FLAT's at short sequence lengths.
On the other hand, \name (+Binding) achieves high utilization across the board and, at long sequence lengths, reaches almost 100\% utilization.
Both baselines achieve slightly higher utilization on XLM, which can be attributed to the higher intensity caused by a larger embedding dimension ($E$/$F$).

Figure~\ref{fig:eval:util:breakdown} explores this phenomenon in more detail, breaking down the utilization by Einsum.
\name effectively hides both the costs of the memory traffic and softmax compute, allowing it to achieve high 2D array utilization while spending most of the cycles on the tensor products.

\textbf{Speedup.}
Figure~\ref{fig:eval:attn:speedup} shows that \name achieves an average speedup of $10\times$ over the unfused baseline and $6.7\times$ over FLAT.
We note \name achieves lower speedup on XLM only because the baselines are able to achieve higher utilization of the 2D array on this transformer (Figure~\ref{fig:eval:util:2d}).

\textbf{Energy.}
Figure~\ref{fig:eval:attn:energy} shows that \name uses $77\%$ the energy of the unfused baseline and $79\%$ the energy of FLAT.\footnote{FLAT reports larger energy savings over the unfused baseline because it only reports energy associated with DRAM traffic during the tensor products.}
The energy use of the unfused baseline and FLAT are dominated by the DRAM access energy, the global buffer access energy, and the $QK$ and $AV$ (Einsums~\ref{eq:attn:basic:qk} and~\ref{eq:attn:basic:av}) compute energy.
\name achieves its energy savings by significantly reducing the DRAM and global buffer access energies.
In fact, $\geq 95\%$ of the energy used by \name across all models and sequence lengths goes to the compute performed by the MACC functional units in the 2D array.

\subsection{Evaluating Transformer Inference}
\label{sec:eval:end2end}

To evaluate the benefits of \name on end-to-end transformer inference, we include the other required linear layers (Section~\ref{sec:background:transformers}).
We use Timeloop to search for optimal mappings for these linear layers and use the same mappings for all three accelerator configurations.
The attention modeling remains the same as Section~\ref{sec:eval:attn}.

\textbf{Speedup.}
Figure~\ref{fig:eval:end2end:speedup} shows the performance improvement achieved by \name.
Across the sequence lengths tested, \name achieves an average speedup of $7.6\times$ over the unfused baseline and $5.3\times$ over FLAT.
As discussed in Section~\ref{sec:background:transformers}, as sequence length grows, attention becomes a larger fraction of the total required compute.
Therefore, at 1M tokens, \name achieves an average $10\times$ speedup over the unfused baseline and $7.5\times$ speedup over FLAT.

\textbf{Energy.}
Figure~\ref{fig:eval:end2end:energy} shows the energy reduction achieved by \name.
Here, we see similar results: as attention becomes a larger fraction of the kernel, the energy reduction increases.
\name uses $82\%$ of the unfused baseline and $83\%$ of FLAT's energy during end-to-end inference.

\begin{figure}[t!]
    \includegraphics[width=0.48\textwidth]{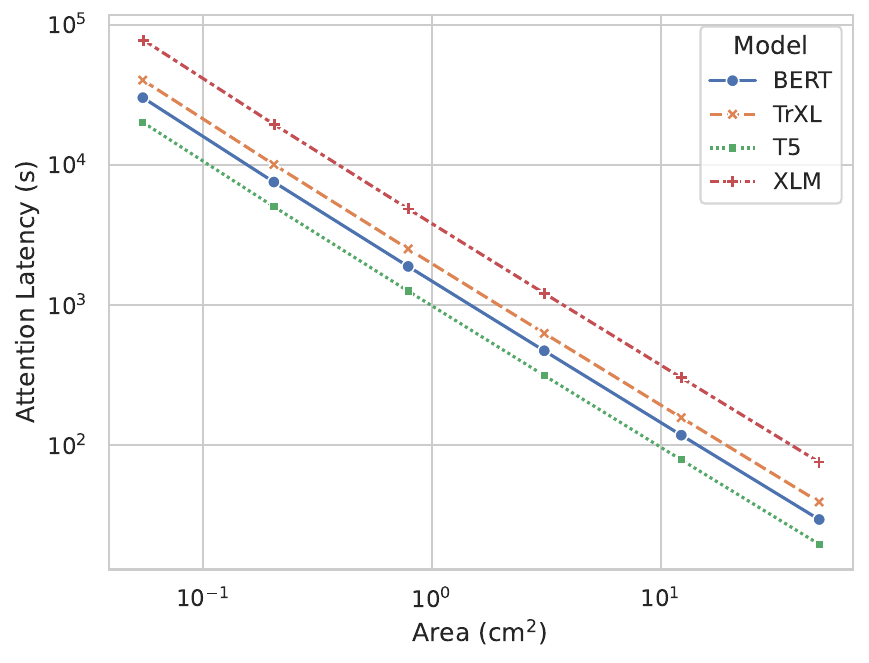}
    \caption{Pareto-optimal curves for \name at sequence length 256K.}
    \label{fig:eval:pareto}
\end{figure}

\subsection{Pareto-Optimality of \name}
We further observe that by varying the size of the PE array (between $16 \times 16$ and $512 \times 512$) and setting the global and per-PE buffers to accommodate the resulting pipelined/interleaved binding, we generate a family of designs for efficient transformer inference.
\section{Related Work}
\label{sec:related}

Spatial architectures have been applied successfully to a variety of domains in academia~\cite{eyeriss, Eyeriss2, plasticine, triginst} and industry~\cite{tpu, trainium}. 
Beyond FLAT~\cite{flat} (discussed in the main body of the paper), TileFlow~\cite{tileflow} is a framework for modeling and searching for efficient fused dataflows (including for attention) on spatial architectures. 
Though TileFlow does explore a broader space of dataflows than FLAT, even implementing the 2-pass softmax cascade (Section~\ref{sec:einsums:2pass}), its dataflows remain softmax-compute limited.
Recent work has explored the scheduling/compilation of a multi-Einsum kernels~\cite{autosa, looptree, tileflow}. 
However, these works explore a limited set of transformations, making \name's inter-Einsum interleaving not discoverable.

Quantization and sparsity have also been successfully applied to reduce the transformer inference compute and live footprint.
We view these schemes as complementary to our work. 
GPTQ~\cite{gptq}, AWQ~\cite{awq}, and LLM.int8()~\cite{llm.int8} quantize model weights to 4 or 8 bits without significant accuracy degradation.
Outlier-aware quantization schemes like GOBO~\cite{gobo} and OliVe~\cite{olive} quantize both weights and activations to a low-bit precision on specific hardware designs. 
SpAtten~\cite{spatten} prunes entire tokens and heads, while
Sanger~\cite{sanger} and DOTA~\cite{dota} use quantized or low-rank projected $Q$ and $K$ tensors to estimate which values of $QK$ and $A$ can be safely pruned. 
All of these algorithms are expressible as cascades of Einsums, and therefore, may be combined with \name to improve performance and energy efficiency, though we leave their specification and implementation to future work.
\section{Conclusion}

This paper advanced the state of the art in spatial accelerator design for transformer inference.
To do so, we expressed attention and its variants as cascades of Einsums.
We used these cascades to reason about attention's characteristics, independent of its mapping/scheduling.
Using these principles, we proposed \name---an accelerator that uses deep fusion and fine-grain pipelining to map attention onto a spatial architecture.
\name achieves $\sim100\%$ utilization of both PE arrays, demonstrating $6.7\times$ speedup over the prior state-of-the-art (FLAT) using $79\%$ of the energy on attention and $5.3\times$ speedup over FLAT using $83\%$ of the energy on end-to-end inference.

Our work shows that cascades of Einsums provide a powerful abstraction for representing and analyzing domain-specific kernels.
Future work may explore their application to other attention variants (e.g., those exploiting quantization and sparsity) or even other domains (e.g., fully homomorphic encryption, scientific computing, relational algebra, etc.).
Doing so enables mapping-agnostic analysis and may elucidate previously undiscovered cascades and schedules for these algorithms.
\section*{Acknowledgment}


We thank 
the anonymous MICRO and MLArchSys reviewers for their feedback on submitted versions of the work;
Abhimanyu Bambhaniya, Sheng-Chun Kao and Tushar Krishna for discussions on FLAT; and finally
Tanner Andrulis and Angshuman Parashar for help with Timeloop.
We would also like to thank 
Nafea Bshara, 
Ron Diamant,
Serina Tan,
Stephen Neuendorffer, 
Yakun Sophia Shao,
Hongbin Zheng, and others at Amazon, AMD, and NVIDIA for helpful discussions about the work as it matured.

\appendix

\section{Artifact Appendix}

\subsection{Abstract}

In this artifact, we provide Timeloop and Accelergy models of the accelerator \name, an accelerator for encoder-style transformer inference. 
For ease-of-use, we provide a Docker container and a set of Jupyter notebooks through which to run the experiments.
This artifact can
be evaluated on an x86-84 machine with 5 GB of disk space.

\subsection{Artifact check-list (meta-information)}

{\small
\begin{itemize}
  \item {\bf Algorithm: }Timeloop/Accelergy model of the \name accelerator and the baselines it was evaluated against
  \item {\bf Program: }Python, Timeloop, Accelergy
  \item {\bf Run-time environment: }Docker, Jupyter
  \item {\bf Hardware: }x86-64 machine
  \item {\bf Output: }Plots generated from scripts
  \item {\bf Experiments: }Modeling of the five different accelerator design points via Timeloop and Accelergy models
  \item {\bf How much disk space required (approximately)?: } 5GB
  \item {\bf How much time is needed to prepare workflow (approximately)?: }20 minutes
  \item {\bf How much time is needed to complete experiments (approximately)?: }9 hours
  \item {\bf Publicly available?: }Yes
  \item {\bf Archived (provide DOI)?: }Provided after evaluation
\end{itemize}
}

\subsection{Description - How to access}

The artifact is hosted on Github at \url{https://github.com/
FPSG-UIUC/micro24-fusemax-artifact}.
Following the instructions in this repository will allow you to install the relevant dependences, run the experiments, and display the graphs.
System requirements can be found at \url{https://github.com/FPSG-UIUC/micro24-fusemax-artifact/blob/main/README.md#system-requirements}.

\subsection{Installation}

Installation instructions can be found at \url{https://github.com/FPSG-UIUC/micro24-fusemax-artifact/blob/main/README.md#installation}.

\subsection{Evaluation}

Evaluation instructions can be found at \url{https://github.com/FPSG-UIUC/micro24-fusemax-artifact/blob/main/README.md#run-experiments}.

\subsection{Expected Results}

Graphs will be displayed within the Jupyter notebook and/or found in \texttt{workspace/outputs/generated/<timestamp or default>/figs/}.
They can be compared with Figures 6-12 in the paper or the corresponding figures in \texttt{workspace/outputs/pregenerated/figs/}.

\subsection{Methodology}

Submission, reviewing and badging methodology:

\begin{itemize}
  \item \url{https://www.acm.org/publications/policies/artifact-review-and-badging-current}
  \item \url{https://cTuning.org/ae}
\end{itemize}



\bibliographystyle{IEEEtranS}
\bibliography{refs, architecture, chris, deep_learning, tensors}

\end{document}